\documentclass[preprint]{elsarticle}

\usepackage{graphicx}
\usepackage{amsmath}
\usepackage{amssymb}
\usepackage{lineno}

\usepackage{epstopdf}
\usepackage[linesnumbered, ruled]{algorithm2e}
\usepackage{nomencl}
\usepackage{bm}
\usepackage{subfigure}
\usepackage[colorlinks=true, pdfstartview=FitV, linkcolor=black, citecolor= black, urlcolor= black]{hyperref}
\usepackage{footnpag}			      	% make footnote symbols restart on each page

\usepackage{wrapfig}

\makenomenclature

\makeindex

\journal{Acta Astronautica}

\begin{document}
	
\begin{frontmatter}
\title{Simultaneous navigation and mascon gravity estimation around small bodies}
\author[1,2]{Julio C. Sanchez\corref{cor1}} 
\ead{jsanchezm@us.es}
\author[2]{Hanspeter Schaub}
\address[1]{Departamento de Ingenieria Aeroespacial y Mecanica de Fluidos, Universidad de Sevilla, 41092, Sevilla, Spain}
\address[2]{Department of Aerospace Engineering Sciences, University of Colorado Boulder, 80309, Boulder, United States}

\cortext[cor1]{Corresponding author}


\begin{abstract}
This manuscript develops a simultaneous navigation and gravity estimation strategy around a small body. The scheme combines dynamical model compensation with a mascon gravity fit. Dynamical compensation adds the unmodeled acceleration to the filter state. Consequently, the navigation filter is able to generate an on-orbit position-unmodeled acceleration dataset. The available measurements correspond to the landmarks-based navigation technique. Accordingly, an on-board camera is able to provide landmark pixels. The aforementioned position-unmodeled acceleration dataset serves to train a mascon gravity model on-board while in flight. The training algorithm finds the optimal mass values and locations using Adam gradient descent. By a careful choice of the mascon variables and constraints projection, the masses are ensured to be positive and within the small body shape. The numerical results provide a comprehensive analysis on the global gravity accuracy for different estimation scenarios.
\end{abstract}
\end{frontmatter}

\section{Introduction}

Small bodies exploration is of great scientific interest because these objects are able to explain Solar System formation processes \cite{Castillo2012}. They could also collide with Earth, thus it is pertinent to design and test asteroid deflection techniques \cite{Cheng2018}. There are several small body missions planned in the 2020s decade. Two prominent examples include Psyche \cite{William2018} which will explore the metallic asteroid of the same name and Hera \cite{Michel2022} which will analyze in-situ the DART \cite{Rivkin2021} impact on Dydimos's moon. Currently, OSIRIS-REx mission \cite{Lauretta2017} is returning to Earth with collected samples from asteroid Bennu.

In the vicinity of a small body, the spacecraft motion may be highly perturbed from a Keplerian orbit. This is due to the strong effects of inhomogeneous gravity and solar radiation pressure \cite{Scheeres2012}. If unaccounted for these perturbations can lead to escape or collision trajectories. Therefore, they have to be taken into account in order to ensure safe flight operations. However, before a small body mission launches, only coarse information of the body shape can be inferred through Earth-based radar and telescope measurements. While this is not a concern for solar radiation pressure, it certainly is for the small body gravity field determination. As a matter of fact, the inhomogeneous gravity is only observable in the close neighborhood of the small body. A practical example is the NEAR mission around the asteroid Eros. The Eros spherical harmonics gravity was observable, relative to their uncertainty, up to 10th degree \cite{Miller2002}. The gravity determination process largely relied on processing Deep Space Network (DSN) range and range-rate with long (up to 30 days) optical observations arcs. Although this methodology reduces the needs for on-board computation, it slows down the mission timeline due to long data arcs, signal delays and DSN constrained accessibility. Consequently, automating the gravity field determination process has the potential to speed up the transition to low altitude operations (with high scientific return). Moreover the dependence on costly and constrained Earth-based infrastructure is reduced.

In the last decade, several publications \cite{Leonard2012, Hesar2015, Stacey2018, Dennison2023, Biggs2019, Sanchez2022, Pasquale2022, Martin2022, Sanchez2023} are proposing autonomous gravity estimation strategies. References \cite{Leonard2012,Hesar2015} revisited NEAR mission orbit determinaton and gravity field estimation by using linked autonomous interplanetary satellite orbit navigation (LiAISON). This technique exploits the gravity field asymmetries along with relative satellite-to-satellite range and range-rate measurements to provide precise orbit determination. In particular, \cite{Hesar2015} demonstrates that Eros gravity field is theoretically observable up to 9th degree by using a suitable beacon-satellite configuration. References \cite{Stacey2018, Dennison2023} focus on close formation satellites (namely swarm) for simultaneous navigation and asteroid characterization. These works fuse landmarks-based and inter-satellite ranging measurements within a centralized unscented Kalman filter (UKF). The UKF estimates the swarm state and spherical harmonics coefficients (amongst other variables of interest). Second-order gravity and solar sail reflection coefficient degradation are inferred in \cite{Biggs2019}. The parameters estimation is done through recursive least-squares fitting of the unmodeled acceleration (which is inferred with an extended state observer). However, the simulations lack realism as perfect knowledge of the spacecraft position is assumed. Reference \cite{Sanchez2022} develops a gravity model learning-based predictive control for orbit-attitude station-keeping. A spherical harmonics gravity model is learned by averaging UKF individual estimates of a satellite constellation (which uses landmarks and laser-based navigation). Multiple satellites are able to mitigate outliers and augment the convergence rate with respect to a single satellite. The results also show the positive impact of model learning in orbit control accuracy and torque demands. Two different machine-learning approaches are used in \cite{Pasquale2022, Martin2022}. Reference \cite{Pasquale2022} utilizes Hopfield neural networks for spherical harmonics gravity estimation. The results show that the neural network estimation is able to provide similar performance as extended Kalman filtering (EKF) with a lower computational cost. However, the navigation component is simplified by assuming zero-mean white Gaussian noise of position and velocity. Differently, \cite{Martin2022} couples a dynamical model compensated filter (which estimates a position-unmodeled acceleration dataset) with a physics-informed neural network gravity representation. While marginal improvement with respect to a point-mass model is achieved, it seems a promising data-driven technique for gravity field determination. 

The autonomous gravity estimation literature shows a clear preference for the spherical harmonics model \cite{Leonard2012, Hesar2015,Stacey2018,Dennison2023,Biggs2019,Sanchez2022,Pasquale2022} (except Reference \cite{Martin2022}). The spherical harmonics model is generally valid only outside the body circumscribing sphere (namely Brillouin sphere). This is of great concern when planning low altitude operations (descent, landing or touch and go) because the gravity prediction may be flawed in that domain. In order to avoid spherical harmonics Brillouin sphere divergence, alternative gravity models \cite{Werner1996,Garmier2001,Russell2012,Gao2019,Martin2022_bis,Izzo2022} (amongst others) can be employed. For on-board execution and estimation, several considerations have to be analyzed (e.g. computational efficiency) besides accuracy. The polyhedron \cite{Werner1996} and novel neural density fields \cite{Izzo2022} models provide a high global accuracy but are expensive computationally. The ellipsoidal harmonics model \cite{Garmier2001} reduces the size of the spherical harmonics divergence domain but does not avert it. The mascon model \cite{Russell2012} can be adjusted to offer a balance between accuracy and computational load. Machine-learning based models (also \cite{Izzo2022}) are recently proposed in the form of Gaussian processes \cite{Gao2019} and physics informed neural networks \cite{Martin2022_bis}. These are a suitable data-driven options but are on an early stage in terms of convergence outside the dataset domain (e.g. lack of data close to the surface).

Under the previous considerations, this work explores the mascon model for autonomous gravity estimation. This model represents the gravity field as the joint contributions of multiple simple shapes (e.g. spheres) discretizing the small body. A typical research direction consists on using mascon distributions that approximate the constant density polyhedron model \cite{Chanut2015}. Alternatively, \cite{Russell2012,Colagrossi2015,Garcia2021} use optimization techniques to find the mascon distribution that fits a dataset. Reference \cite{Russell2012} fits the Earth's gravity field beyond $J_2$ by assuming the point masses are fixed a-priori. In \cite{Colagrossi2015}, a genetic algorithm finds the optimal point masses locations and its values. In a similar spirit, \cite{Garcia2021} employs a Newton-Raphson algorithm to compute the optimal distribution of a low number of point masses. The previous works assume true position-acceleration (or potential in \cite{Russell2012}) datasets for their mascon trainings. Moreover, \cite{Russell2012,Colagrossi2015} datasets contain abundant and homogeneously distributed samples across the spatial domain while \cite{Garcia2021} relies on low altitude points. While these conditions are acceptable for gravity field modeling analysis, they are possibly unfeasible during the on-orbit scenario. Specifically, the data may be highly concentrated on a specific orbital regime (though low altitude samples may be obtained with additional hardware \cite{Hockman2022} or ejecta observations). Even more crucial, the dataset is also corrupted by navigation uncertainties. Following the previous consideration, the on-orbit generation of the gravity acceleration data poses a question on itself. To this end, the concept of dynamical model compensation (DMC) could be tailored to this specific application. The DMC key idea is to estimate the unmodeled dynamics component (as an acceleration) along with the spacecraft state. Operating as such, the persistent unknown dynamical bias is removed from the filter process which is of high interest for precise orbit determination \cite{Myers1975,Leonard2013}. 

This manuscript combines a dynamical model compensated unscented Kalman filter (DMC-UKF) with the training of a mascon gravity model. Following previous works in small body navigation \cite{Leonard2012,Hesar2015,Vetrisano2016,Stacey2018,Zhu2019,Sanchez2022}, the spacecraft is assumed to be equipped with a camera that tracks landmarks on the small body surface. The identified landmark pixels on the image plane (which is out of the scope of this work) constitute the DMC-UKF measurements. The DMC-UKF generates an on-orbit position-unmodeled acceleration dataset over time. Then, this dataset serves to fit a mascon gravity model using Adam gradient descent. This paper extends and completes the authors initial communication of \cite{Sanchez2022}. The major breakthrough is that the mascon training has now the feature to find the optimal masses spatial distribution (instead of just the masses values). Moreover, physical constraints such as masses positiveness and its containment within the body shape are explicitly accounted for. To resume, the main contributions of this work are: 1) the autonomous gravity estimation of a mascon model which does not diverge within Brillouin sphere; 2) the resulting mascon distribution is physically consistent (positive masses within the body shape) and, 3) the potential flexibility (to other gravity models) of the proposed strategy since the dataset generation is decoupled from gravity estimation. 

The structure of the paper is as follows. Section II describes spacecraft dynamics around a small body. Section III presents the camera model and DMC-UKF algorithm. Section IV develops the mascon gravity training algorithm and its integration with DMC-UKF. Section V shows numerical results of interest. Section VI concludes this paper and states future research directions.

\section{Dynamics around a small body}

\begin{figure}[] 
	\begin{center}
		\includegraphics[width=11cm,height=11cm,keepaspectratio,,trim={0 0cm 0 0cm},clip]{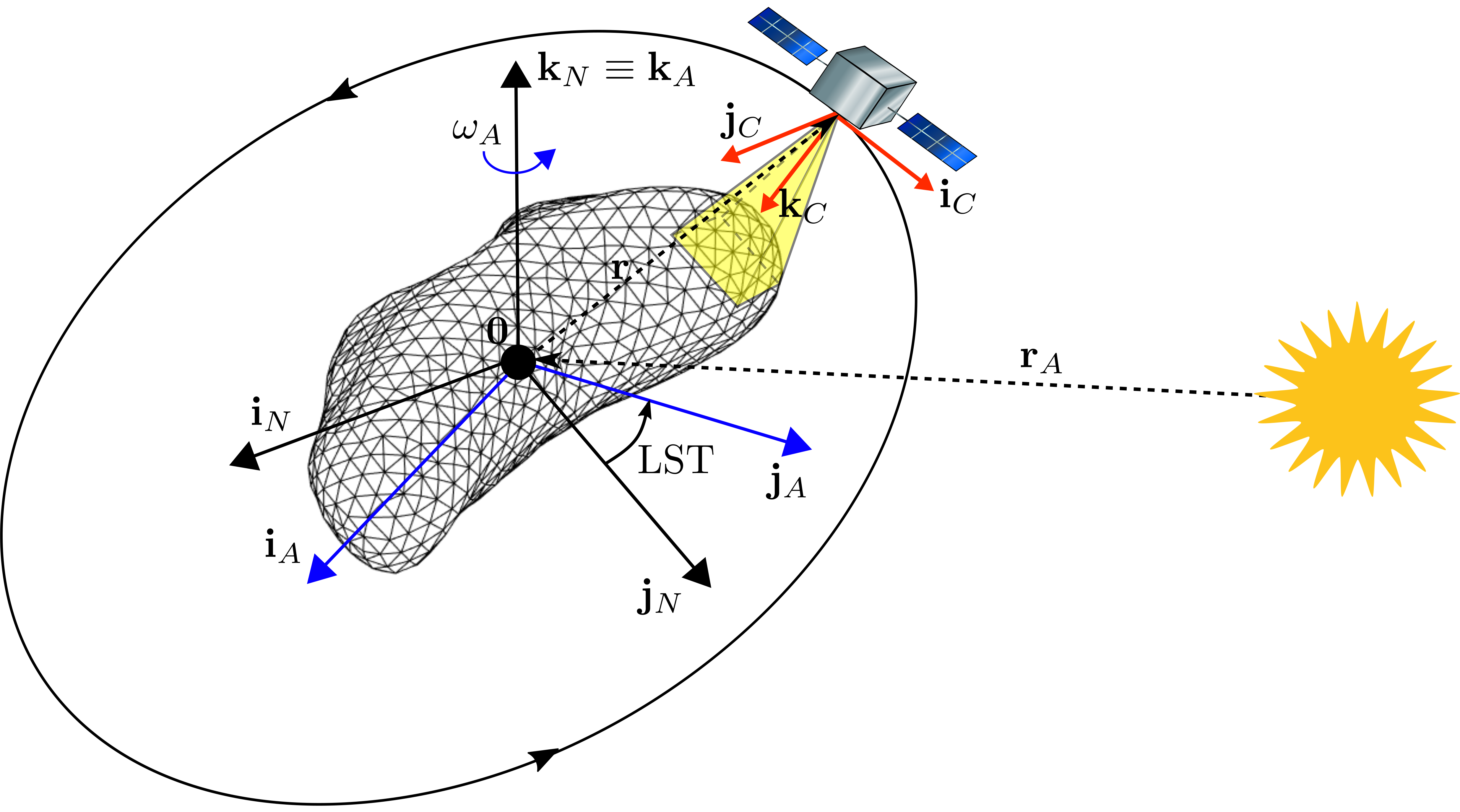}
	\end{center}
	\vspace{-0.5cm}
	\caption{Reference frames.}	
	\label{fig:referenceframe}
\end{figure}

A spacecraft in the proximity of a small body is perturbed by the gravity field, the Sun's third body gravity and the solar radiation pressure as
\begin{equation}
	\ddot{\mathbf{r}}^N=(\mathbf{R}^{A}_{N})^T\mathbf{a}_{\text{poly}}+\mathbf{a}^N_{\odot}+\mathbf{a}^N_{\text{SRP}},\label{eq:spacecraft_dynamics}
\end{equation}
where $\mathbf{r}^N$ is the spacecraft position expressed in an inertial small body centred frame $N\equiv\{\mathbf{0}:\mathbf{i}_N,\mathbf{j}_N,\mathbf{k}_N\}$ where $\mathbf{0}$ is the small body center of mass and $\mathbf{k}_N$ its major inertia axis. The Sun's third body gravity perturbation is  $\mathbf{a}^N_{\odot}$ and the solar radiation pressure acceleration is $\mathbf{a}^N_{\text{SRP}}$. The term $\mathbf{a}_{\text{poly}}$ is the small body polyhedron gravity acceleration \cite{Werner1996} which is assumed as the ground truth (it is further detailed in paragraph~\ref{sec:polyhedron}). This vector is expressed in a rotating small body centred frame $A\equiv\{\mathbf{0}:\mathbf{i}_A,\mathbf{j}_A,\mathbf{k}_A\}$. Let consider $\mathbf{k}_A\equiv\mathbf{k}_N$, thus the equatorial plane is defined by $\mathbf{i}_A\mathbf{j}_A$. In order to ease the notation, the superscript $A$ is omitted for any vector expressed in the rotating frame. The $N$ and $A$ frames are related through the direction cosine matrix $\mathbf{R}^{A}_{N}$. By assuming the small body rotates at a constant rate $\omega_A$ around its major inertia axis (as is the usual case for small bodies), the direction cosine matrix $\mathbf{R}^{A}_{N}$ is 
\begin{equation}
	\mathbf{R}^{A}_{N}=\begin{bmatrix}
		\cos{(\text{LST}_0+\omega_A t)} & \sin{(\text{LST}_0+\omega_A t)} & 0\\
		-\sin{(\text{LST}_0+\omega_A t)} & \cos{(\text{LST}_0+\omega_A t)} & 0\\
		0 & 0 & 1\\
	\end{bmatrix},\label{eq:rotmatrix}
\end{equation}
where $\text{LST}_0$ is the small body initial local sidereal time and $t$ is the elapsed simulation time. The reference frames are illustrated in Fig.~\ref{fig:referenceframe}. 

\subsection{Solar perturbations}

The Sun's third body gravity $\mathbf{a}^N_{\odot}$ and the solar radiation pressure $\mathbf{a}^N_{\text{SRP}}$, are described as
\begin{equation}
	\mathbf{a}^N_{\odot}=-\mu_{\odot}\left(\frac{\mathbf{r}^N_{A}+\mathbf{r}^N}{\lVert\mathbf{r}^N_{A}+\mathbf{r}^N\rVert^3}-\frac{\mathbf{r}^N_{A}}{r^3_A}\right),\quad \mathbf{a}^N_{\textup{SRP}}=\frac{C_RSW_{\bigoplus}r^2_{\bigoplus}}{mc\lVert \mathbf{r}^N_{A}+\mathbf{r}^N\rVert^3}(\mathbf{r}^N_{A}+\mathbf{r}^N),\label{eq:solar_perturbations}
\end{equation}
where $\mathbf{r}^N_{A}$ is the small body relative position with respect to the Sun and $\mu_{\oplus}=1.3271244\cdot10^{20}~\textup{m}^{3}/\text{s}^2$ is the Sun's standard gravity parameter. The expression of the solar radiation pressure corresponds to the simple cannonball model. The term $m$ is the spacecraft mass, $C_R$ its reflection coefficient, $S$ is the exposed surface to the photons, $W_{\bigoplus}=1366~\text{W}/\text{m}^2$ is the mean energy flux received from the Sun at the mean orbital distance of $r_{\bigoplus}=1~\text{AU}$  and $c=3\cdot10^{8}~\text{m}/\text{s}^2$ is the speed of light.

\subsection{Small body gravity field}

Throughout this paper, two models are used to characterize the small body gravity. As previously mentioned, the polyhedron model is assumed as ground truth (for validation) while mascon models are utilized for gravity estimation. 

\subsubsection{Polyhedron model}\label{sec:polyhedron}

Reference \cite{Werner1996} expresses the exterior gravity field generated by a constant density polyhedron (described by faces and vertexes) as
\begin{equation}
	\mathbf{a}_{\text{poly}}=-\frac{\mu}{V}\left(\sum_{e\in\text{edges}}\mathbf{E}_e\cdot\mathbf{r}_eL_e-\sum_{f\in\text{faces}}\mathbf{F}_f\mathbf{r}_fw_f\right),\label{eq:polyhedron_model}
\end{equation}
where $V$ is the body volume, $\mathbf{r}_e$ is the relative position of the evaluation point with respect to the edge origin, $\mathbf{E}_e$ is the dyad product resulting from the edge and face normals, $L_e$ is the potential of the edge as a 1D wire, $\mathbf{r}_f$ is the relative position of the evaluation point with respect to a vertex on a face, $\mathbf{F}_f$ is the outer product of the face normal vector and $w_f$ is the solid angle of the face as viewed from the evaluation point. The explicit details of the derivation can be consulted in Ref.~\cite{Werner1996}. This model is useful for validation because it is globally accurate for the given shape with constant density. Although the constant density assumption may not hold in a real scenario, it is considered acceptable for synthetic simulations.  

\subsubsection{Mascon model} 

The mascon model characterizes the gravity field by adding the contributions of several simple volume elements (namely mascons). These can be homogeneous spheres with different densities. In this work only the exterior (outside the body volume) component of the gravity field is of interest, thus the mascon elements are denoted as point masses. Then, the mascon gravity is
\begin{equation}
	\mathbf{a}_{M}=-\sum^{n}_{k=0}\mu_{M_k}\frac{\mathbf{r}-\mathbf{r}_{M_k}}{\lVert\mathbf{r}-\mathbf{r}_{M_k}\rVert^3},\label{eq:mascon_model}
\end{equation}
where $n+1$ is the number of point masses while $\mu_{M_k}$ and $\mathbf{r}_{M_k}$ are, respectively, the standard gravity parameter and position of each one. In order to avoid singularities in the external gravity field evaluation, the point masses shall be within the small body shape. 

\section{Dynamical model compensated UKF}\label{sec:DMC-UKF}

This section starts describing the pinhole camera model which maps 3D points to pixels in the image plane. Then, the dynamical model compensated unscented Kalman filter is stated. The DMC-UKF has the task of generating a position-unmodeled acceleration dataset for gravity estimation.

\subsection{Pinhole camera model}

\begin{figure}[] 
	\begin{center}
		\includegraphics[width=11cm,height=11cm,keepaspectratio,,trim={1cm 2.05cm 0 0.4cm},clip]{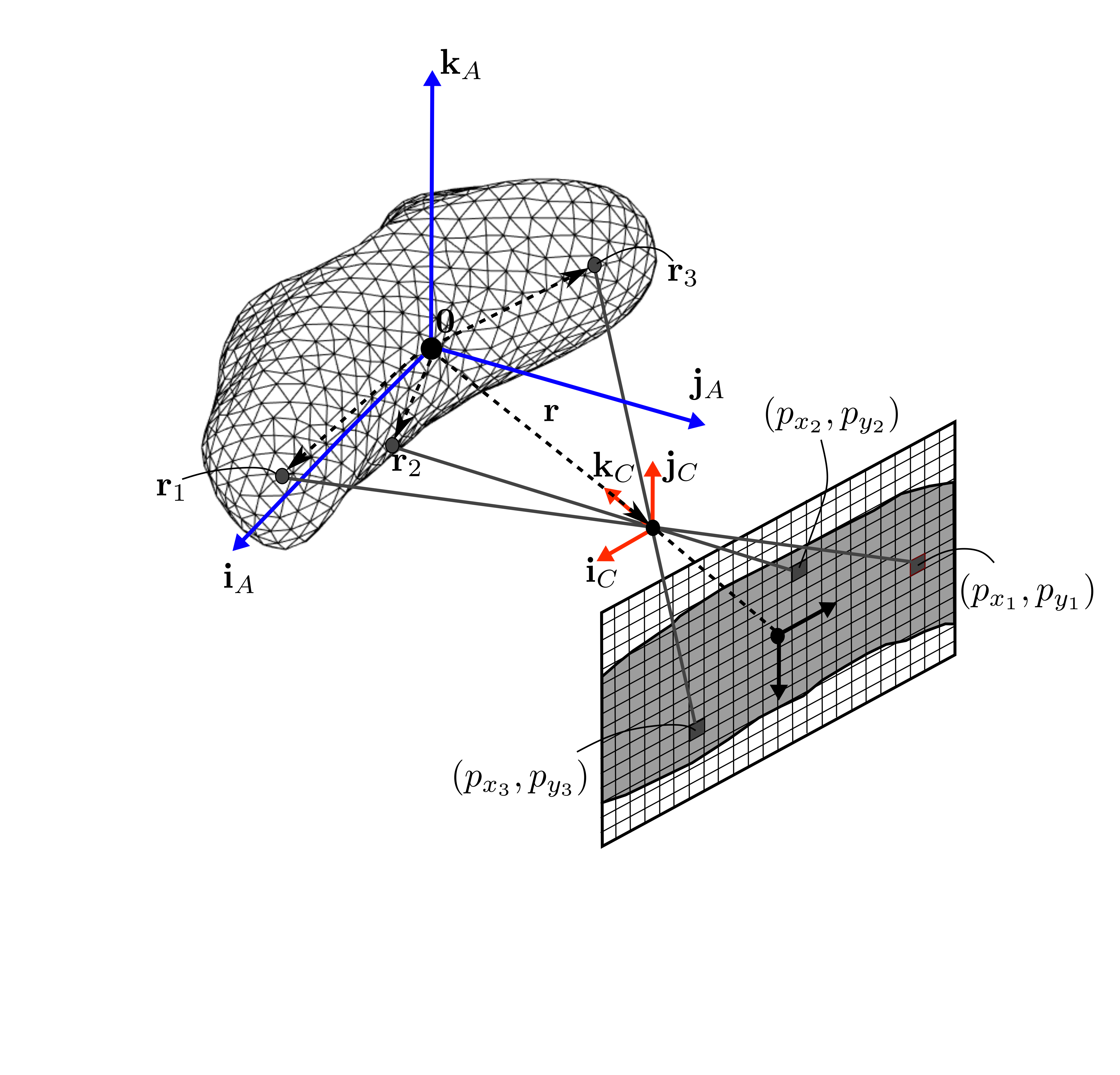}
	\end{center}
	\vspace{-0.5cm}
	\caption{Illustration of the pinhole camera model.}	
	\label{fig:camerapinhole_model}
\end{figure}

On-board optical navigation is a state of the art technique to acquire relative measurements with respect to a small body. From a high-level perspective, the underlying process can be simulated using a pinhole camera model. This model provides algebraic expressions that map 3D points to pixels. Let define the camera reference frame as $C\equiv\{\mathbf{r}:\mathbf{i}_C,\mathbf{j}_C,\mathbf{k}_C\}$ which, for simplicity, is centred at the spacecraft center of mass $\mathbf{r}$ (since the offset between camera aperture and center of mass is a known constant) and $\mathbf{k}_C$ is the optical axis pointing towards the viewing direction (see Fig.~\ref{fig:referenceframe}). Then, the projection of a 3D point $(x_C,y_C,z_C)$, expressed in the $C$ frame, to virtual image plane coordinates $(u,v)$ is as follows
\begin{equation}
	\begin{bmatrix}
		u\\
		v\\
	\end{bmatrix}=\frac{f}{z_C}
	\begin{bmatrix}
		x_C\\
		y_C\\
	\end{bmatrix},\label{eq:cam1}
\end{equation} 
where $f$ is the camera focal length. Since the image is digital, the 3D point maps to a pixel $(p_x,p_y)$ as
\begin{equation*}
	p_x = \begin{cases}
		\text{ceil}(u/w_p)-0.5 & \text{if}\quad u \geq 0,\\
		\text{floor}(u/w_p)+0.5 & \text{if}\quad u < 0,
	\end{cases} \quad
	p_y = \begin{cases}
		\text{ceil}(v/w_p)-0.5 & \text{if}\quad v \geq 0,\\
		\text{floor}(v/w_p)+0.5 & \text{if}\quad v < 0.
	\end{cases}\label{eq:cam2}
\end{equation*} 
Although the pixel should be an integer variable, its center is taken as the practical measurement in order to reduce numerical dispersion. The term $w_p$ is the pixel width which is determined by the camera sensor size and its resolution. An important camera parameter is the field of view $\text{FOV}$ which quantifies how much is visible through the lens. The FOV is characterized by horizontal and vertical angles related to the focal length and sensor size as
\begin{equation}
	\text{FOV}\equiv2\arctan(n_{p_x}w_p/2f) \times 2\arctan(n_{p_y}w_p/2f),\label{eq:focallenght}
\end{equation}
where $n_{p_x}$ and $n_{p_y}$ are, respectively, the horizontal and vertical number of pixels. In Eq.~\eqref{eq:focallenght}, the focal length $f$ is the varying parameter that controls the $\text{FOV}$ size. The camera pinhole model geometry is illustrated in Fig.~\ref{fig:camerapinhole_model}. 

\subsection{DMC-UKF}

The DMC-UKF jointly estimates the spacecraft state (position and velocity) and the unmodeled perturbing acceleration. The dynamical model compensation concept directly tracks the unknown dynamics component through estimation of the unmodeled acceleration signal \cite{Myers1975, Leonard2013}. In this work, the previous concept is embedded in an unscented Kalman filter \cite{Wan2000}. The UKF is a suboptimal non-linear filter for Gaussian distributions. Its main feature is the unscented transform (UT) which approximates the result of applying a non-linear function to a Gaussian distribution. To this end, the UT creates a symmetrical set of samples (namely sigma points $\pmb{\chi}_{\mathbf{x}}$) around the mean $\hat{\mathbf{x}}$ of the initial Gaussian distribution $\mathbf{x}\sim N_{n_x}(\hat{\mathbf{x}},\pmb{\Sigma}_{\mathbf{x}\mathbf{x}})$ as
\begin{equation}
	\pmb{\chi}^{[i]}_{\mathbf{x}}=\hat{\mathbf{x}}+\text{sgn}(i)\cdot\left(\sqrt{(n_x+\lambda)\pmb{\Sigma}_{\mathbf{xx}}}\right)_{|i|},\quad i=-n_x\hdots0\hdots n_x.\label{eq:ut1}
\end{equation}
Note that $N_{n_x}$ denotes a Gaussian distribution of dimension $n_x$ .Then, in order to approximate the distribution $\mathbf{y}=\mathbf{f}(\mathbf{x})$, the UT passes each sigma point through the function $\mathbf{f}$. The sigma points outcome reconstructs the final distribution in a Gaussian form $\mathbf{y}\sim N_{n_y}(\mathbf{y},\pmb{\Sigma}_{\mathbf{y}\mathbf{y}})$ as
\begin{equation}
	\hat{\mathbf{y}}=\sum^{n_x}_{i=-n_x}w^{[i]}_m\mathbf{f}(\pmb{\chi}^{[i]}_{\mathbf{x}}),\quad\quad \pmb{\Sigma}_{\mathbf{y}\mathbf{y}}=\sum^{n_x}_{i=-n_x}w^{[i]}_{c}(\hat{\mathbf{y}}-\mathbf{f}(\pmb{\chi}^{[i]}_{\mathbf{x}}))(\hat{\mathbf{y}}-\mathbf{f}(\pmb{\chi}^{[i]}_{\mathbf{x}}))^T,\label{eq:ut2}
\end{equation}
where, following \cite{Wan2000}, the weights are defined as $w_m^{[0]}=\lambda/(\lambda+n_x)$, $w_c^{[0]}=w_m^{[0]}+1-\alpha^2+\beta$ and $w_m^{[i]}=w_c^{[i]}=1/(2n_x+2\lambda),\>i\neq0$. The variables  $\{\alpha,\beta,\lambda\}$ are tuning parameters controlling the spread of sigma points (see Eq.~\eqref{eq:ut1}) and weights. Subsequently, the process propagation, state to measurements mapping and the complete DMC-UKF algorithm are presented.

\subsubsection{Process propagation} 

Let express the DMC-UKF state in the inertial small body centred frame $N$ as  $\mathbf{x}=[(\mathbf{r}^N)^T,(\mathbf{v}^N)^T,(\mathbf{a}^N)^T]^T$. The term $\mathbf{a}^N$ is the unmodeled acceleration acting on the spacecraft. Accordingly, the DMC-UKF process dynamics is 
\begin{equation}
	\frac{d}{dt}\begin{bmatrix}
		\mathbf{r}^N\\
		\mathbf{v}^N\\
		\mathbf{a}^N\\
	\end{bmatrix}=
	\begin{bmatrix}
		\mathbf{v}^N\\
		(\mathbf{R}^{A}_{N})^T\mathbf{a}_{M}(\mathbf{r})+\mathbf{a}^N_{\odot}(\mathbf{r})+\mathbf{a}^N_{\text{SRP}}(\mathbf{r})+\mathbf{a}^N\\
		\mathbf{0}\\	
	\end{bmatrix},\label{eq:process_dynamics}
\end{equation}
where the Sun's third-body gravity and solar radiation pressure parameters are assumed to be known. Accordingly, the unknown dynamics corresponds to the small body gravity (unmodelled components in $\mathbf{a}_{M}$) which is compensated by the unmodeled acceleration $\mathbf{a}^N$. Its time variation is assumed as null ($\dot{\mathbf{a}}^N=\mathbf{0}$), thus representing a zeroth-order Gauss Markov process (white noise). This implies that the measurements sampling rate should be fast enough in order to keep track of the inhomogeneous gravity signal. Other works in the literature have embedded first \cite{Myers1975} and second-order \cite{Leonard2013} Gauss Markov processes. However, the scope of these works is different from the present one. They indirectly account for the unmodeled effects through the acceleration signal while here uncertainty is directly tackled by updating the gravity model. 

The outcome of the process is the a-priori DMC-UKF state which is obtained by propagating Eq.~\eqref{eq:process_dynamics} using a simple forward Euler integration rule
\begin{equation}
	\mathbf{x}^{-}=\mathbf{x}_0+\sum^{N_{\text{int}}-1}_{k=0}\dot{\mathbf{x}}(t_0+k\Delta t_{\text{int}})\Delta t_{\text{int}},\label{eq:process_propagation}
\end{equation}
where $N_{\text{int}}$ is the number of integration points and $\Delta t_{\text{int}}$ the time step.

\subsubsection{State to measurement mapping} 

The available measurement $\mathbf{z}$ comprises the $n_L$ visible landmarks pixels at the observation epoch
\begin{equation}
\mathbf{z}=[p_{x_1},p_{y_1},\hdots,p_{x_{n_L}},p_{y_{n_L}}]^T.
\end{equation}
In order to map the DMC-UKF state to the measurement, the relative positions (expressed in the camera frame $C$) between spacecraft and visible landmark is computed as 
\begin{equation}
\Delta\mathbf{r}^C_l=\mathbf{R}^{C}_{A}(\mathbf{r}_l-\mathbf{R}^{A}_{N}\mathbf{r}^N),\label{eq:map1}
\end{equation}
where $\mathbf{R}^{A}_{N}$ is the direction cosine matrix from inertial to the rotating small body centred frame (see Eq.~\eqref{eq:rotmatrix}). The direction cosine matrix $\mathbf{R}^{C}_{A}$ represents the orientation of the camera with respect to the rotating small body frame. Because this work does not account for attitude determination, the matrix $\mathbf{R}^{C}_{A}$ is assumed to be perfectly known (provided by the attitude determination system). Finally, each landmark pixel can be mapped as
\begin{equation}
\begin{bmatrix}
p_{x_l}\\
p_{y_l}\\
\end{bmatrix}=
\dfrac{f}{w_p}\begin{bmatrix}
\Delta x^C_l/\Delta z^C_l\\
\Delta y^C_l/\Delta z^C_l\\
\end{bmatrix}.\label{eq:map2}
\end{equation}
where the pixel is not discretized in order to better quantify the difference with respect to the actual discrete measurement. 

\subsubsection{DMC-UKF algorithm} 

The implementation of the DMC-UKF step follows \cite{Wan2000} and is described in Algorithm \ref{alg:UKF_pseudocode}. Let consider the current DMC-UKF state Gaussian distribution $\mathbf{x}_0\sim N_9(\hat{\mathbf{x}}_0,\pmb{\Sigma}_{\mathbf{x}\mathbf{x}_0})$. Let also compress the Eq.~\eqref{eq:process_propagation} process integration and Eq.~\eqref{eq:map1}-\eqref{eq:map2} state to measurement map as $\mathbf{f}$ and $\mathbf{g}$ respectively. When measurements are available, the DMC-UKF state is updated as follows. Step 2 computes the a-priori state estimate $\mathbf{x}^{-}\sim N_9(\hat{\mathbf{x}}^{-},\pmb{\Sigma}^{-}_{\mathbf{x}\mathbf{x}})$ by doing the UT of the initial distribution with the process. Step 3 inflates the a-priori state uncertainty with the process noise covariance $\pmb{\Sigma}_{\mathbf{f}\mathbf{f}}$ (it quantifies the mismatch between truth and process dynamics). This uncertainty is a tuning parameter of the algorithm. Step 4 computes an a-priori measurement distribution $\mathbf{z}^{-} \sim N_3(\hat{\mathbf{z}}^{-},\pmb{\Sigma}^{-}_{\mathbf{z}\mathbf{z}})$ by mapping the a-priori state into measurement space with the UT. Step 5 inflates the a-priori measurement covariance with the covariance matrix $\pmb{\Sigma}_{\mathbf{z}\mathbf{z}}$. This matrix accounts for measurement noise. The step 6 computes the cross-correlation matrix $\mathbf{H}_{\mathbf{x}\mathbf{z}}$ between state and measurement. The cross-correlation matrix $\mathbf{H}_{\mathbf{x}\mathbf{z}}$ is used in step 7 to compute the Kalman gain $\mathbf{K}$. Finally, the incoming measurement $\mathbf{z}$ is used in the Kalman update linear equation (see step 8) to compute the a-posteriori DMC-UKF state distribution $\mathbf{x}\sim N_9(\hat{\mathbf{x}},\pmb{\Sigma}_{\mathbf{x}\mathbf{x}})$. This procedure is sequentially repeated each time measurements are available.
\begin{algorithm}[]\caption{DMC-UKF step}\label{alg:UKF_pseudocode}
	\Begin{Apply the UT (see Eq.~\eqref{eq:ut1}-\eqref{eq:ut2}) to Eq.~\eqref{eq:process_propagation} process: $\mathbf{x}^{-}\sim N_9(\hat{\mathbf{x}}^{-},\pmb{\Sigma}^{-}_{\mathbf{x}\mathbf{x}})\equiv\mathbf{f}(N_9(\hat{\mathbf{x}}_0,\pmb{\Sigma}_{\mathbf{x}\mathbf{x}_0}))$\;
		Add the process uncertainty: $\pmb{\Sigma}^{-}_{\mathbf{x}\mathbf{x}}\leftarrow\pmb{\Sigma}^{-}_{\mathbf{x}\mathbf{x}}+\pmb{\Sigma}_{\mathbf{f}\mathbf{f}}$\;
		Apply the UT (see Eq.~\eqref{eq:ut1}-\eqref{eq:ut2}) to Eq.~\eqref{eq:map1}-\eqref{eq:map2} state to measurement mapping: $\mathbf{z}^{-} \sim N_3(\hat{\mathbf{z}}^{-},\pmb{\Sigma}^{-}_{\mathbf{z}\mathbf{z}})\equiv\mathbf{g}(N_9(\hat{\mathbf{x}}^{-},\pmb{\Sigma}^{-}_{\mathbf{x}\mathbf{x}}))$\; 
		Add the measurement uncertainty: $\pmb{\Sigma}^{-}_{\mathbf{z}\mathbf{z}}\leftarrow\pmb{\Sigma}^{-}_{\mathbf{z}\mathbf{z}}+\pmb{\Sigma}_{\mathbf{z}\mathbf{z}}$\;
		Compute the cross-correlation matrix between state and measurements:
		$\mathbf{H}_{\mathbf{x}\mathbf{z}}=\sum\limits^{9}_{i=-9}w_c^{[i]}\left(\pmb{\chi}^{[i]}_{\mathbf{x}}-\hat{\mathbf{x}}^{-}\right)\left(\pmb{\chi}^{[i]}_{\mathbf{z}}-\hat{\mathbf{z}}^{-}\right)^T$\;
		Compute the Kalman gain: $\mathbf{K}=\mathbf{H}_{\mathbf{x}\mathbf{z}}\pmb{\Sigma}^{-1}_{\mathbf{z}\mathbf{z}}$\;
		Update the state with incoming measurements from Eq.~\eqref{eq:position_determination}:
		$\mathbf{x}\sim N_9(\hat{\mathbf{x}},\pmb{\Sigma}_{\mathbf{x}\mathbf{x}})\equiv N_9(\hat{\mathbf{x}}^{-}+\mathbf{K}(\mathbf{z}-\hat{\mathbf{z}}^{-}),\pmb{\Sigma}^{-}_{\mathbf{x}\mathbf{x}}(\mathbf{I}-\mathbf{H}_{\mathbf{x}\mathbf{z}}\mathbf{K}^T))$\;
	}
\end{algorithm}

\section{Mascon gravity estimation}\label{sec:gravity}

The mascon gravity model estimation process is described in this section. The goal is to minimize the mean squared percent error (MSE) of the fitted model with respect to the acceleration dataset. This should be achieved while taking into account mascon physical constraints. To do so, Adam gradient descent is combined with a constraints projection step. 

\subsection{Gravity fitting problem}

This paragraph begins by defining a loss function based on the available dataset.  Then, the mascon model configuration and the constraints projection step are described. 

\subsubsection{Dataset and loss function}

The mascon model is fitted using the on-orbit position-acceleration dataset generated by the DMC-UKF. Let stack the DMC-UKF estimates in the vectors $\mathbf{r}^{\text{data}}_{\mathbf{S}}$ and $\mathbf{a}^{\text{data}}_{\mathbf{S}}$ as
\begin{equation}
	\mathbf{r}^{\text{data}}_{\mathbf{S}}=\begin{bmatrix}\hat{\mathbf{r}}_1\\
		\vdots\\
		\hat{\mathbf{r}}_m\\
	\end{bmatrix},\quad
	\mathbf{a}^{\text{data}}_{\mathbf{S}}=\begin{bmatrix}\mathbf{a}_{M}(\hat{\mathbf{r}}_1)+\hat{\mathbf{a}}_1\\
		\vdots\\
		\mathbf{a}_{M}(\hat{\mathbf{r}}_m)+\hat{\mathbf{a}}_m\\
	\end{bmatrix}.\label{eq:dataset}
\end{equation}
The position and unmodeled acceleration are rotated from the inertial to the rotating small body frame as $\hat{\mathbf{r}}_j=\mathbf{R}^{A}_N(t_j)\hat{\mathbf{r}}^N_j$ and $\hat{\mathbf{a}}_j=\mathbf{R}^{A}_N(t_j)\hat{\mathbf{a}}^N_j$. The acceleration dataset comprises the sum of the unmodeled one with the current mascon model prediction at the position estimate. In order to ease the notation, let use $\mathbf{r}^{\text{data}}_{\mathbf{S}}=[\mathbf{r}^T_1,\hdots,\mathbf{r}^T_m]^T$ and $\mathbf{a}^{\text{data}}_{\mathbf{S}}=[\mathbf{a}^T_1,\hdots,\mathbf{a}^T_m]^T$ from now on to denote the individual components of the dataset. 

The loss function $L$ is to be defined in terms of the discrepancy between the mascon prediction and the dataset. In \cite{Martin2023}, it is highlighted that the use of percent errors (instead of absolute residuals) helps to regularize the gravity solution (as samples are accounted for relative to their magnitude). Following that logic, let define the gravity percent error $\delta a$ as
\begin{equation}
\delta a_j = \dfrac{\rVert\mathbf{a}_M(\mathbf{r}_j)-\mathbf{a}_j\lVert}{\lVert \mathbf{a}_j \rVert},\label{eq:grav_error}
\end{equation}
where $\mathbf{a}_M$ is the mascon model prediction. Then, a straightforward choice of loss function is the mean squared gravity percent errors
\begin{equation}
L=\frac{1}{m}\sum^{m}_{j=1}\delta a^2_j.\label{eq:loss}
\end{equation}

\subsubsection{Mascon model setup}

The mascon distribution should fulfill two basic physical constraints. The masses shall be enclosed within the body shape (interior constraint) and their values should be positive. The first one avoids singularities in the exterior gravity field evaluation. The second ensures that the gravity field is always attractive, thus avoiding local repulsive regions near negative masses (as it is the case in the preliminary work \cite{Sanchez2023}). Although these conditions may seem simple, they are not neccesarily guaranteed in a numerical optimization framework. Another additional condition emerges if the total mass is to be known (as it is assumed in this work). 

It is more convenient to implicitly encode constraints within the mascon model if possible (rather than explicitly tackling them in the optimizer). In particular, the masses positiveness and total mass can be directly enforced by choosing adequate decision variables. The consistency of the mascon distribution with the total mass is ensured by clearing one of the mascon masses from the optimization as
\begin{equation}
\mu_{M_0}=\mu-\sum^{n}_{k=1}\mu_{M_k}.\label{eq:total_mass}
\end{equation} 
The masses positiveness is encoded by changing the decision variable from the gravity parameter $\mu_{M_k}$ to its square-root $\sqrt{\mu_{M_k}}$. It should be noted that this not guarantees $\mu_{M_0}\geq0$ as it is cleared from the optimization (see Eq.~\eqref{eq:total_mass}). This issue has to be tackled within the mascon fitting algorithm along with the interior constraint.  

According to the previous development, the mascon distribution variables can be stacked in the vector $\mathbf{y}_{\mathbf{S}_M}$ as
\begin{equation}
\mathbf{y}_{\mathbf{S}_M}=\begin{bmatrix}
\sqrt{\pmb{\mu}}_{\mathbf{S}_M}\\
\mathbf{r}_{\mathbf{S}_M}\\
\end{bmatrix},\quad \sqrt{\pmb{\mu}}_{\mathbf{S}_M}=\begin{bmatrix}
\sqrt{\mu_{M_1}}\\
\vdots\\
\sqrt{\mu_{M_n}}\\
\end{bmatrix},\quad \mathbf{r}_{\mathbf{S}_M}=\begin{bmatrix}
\mathbf{r}_{M_1}\\
\vdots\\
\mathbf{r}_{M_n}\\
\end{bmatrix},
\end{equation}
where $\sqrt{\pmb{\mu}}_{\mathbf{S}_M}$ and $\mathbf{r}_{\mathbf{S}_M}$ stack the $n$ square-root masses and positions of the distribution. The position of the 0th mass $\mathbf{r}_{M_0}$ is also cleared from the optimization, thus being a fixed parameter. A convenient placement may be $\mathbf{r}_{M_0}=\mathbf{0}$ (which is the one used). Using the previous stacked variables, a compact prediction of the dataset is obtained as
\begin{equation}
\mathbf{A}_{M}(\sqrt{\pmb{\mu}}_{\mathbf{S}_M})^2+\mathbf{a}_{\mathbf{S}_{M_0}}=\mathbf{a}_{\mathbf{S}_M},
\end{equation}
where the squared vector denotes the element-wise product $(\sqrt{\pmb{\mu}}_{\mathbf{S}_M})^2=\sqrt{\pmb{\mu}}_{\mathbf{S}_M}\odot\sqrt{\pmb{\mu}}_{\mathbf{S}_M}$. The matrix $\mathbf{A}_{M}$ exclusively depends on the relative distance between the evaluation points and each mass. The vector  $\mathbf{a}_{\mathbf{S}_{M_0}}$ is an offset arising from clearing the 0th mass and assuming its location fixed. The terms $\mathbf{A}_M$ and $\mathbf{a}_{\mathbf{S}_{M_0}}$ are 
\begin{equation*}
	\mathbf{A}_{M}=\begin{bmatrix}
		-\dfrac{\Delta\mathbf{r}_{11}}{\Delta r^3_{11}}+\dfrac{\Delta\mathbf{r}_{10}}{\Delta r^3_{10}} & \hdots &  -\dfrac{\Delta\mathbf{r}_{1n}}{\Delta r^3_{1n}}+\dfrac{\Delta\mathbf{r}_{10}}{\Delta r^3_{10}} \\
		\vdots & \ddots & \vdots\\
		-\dfrac{\Delta\mathbf{r}_{m1}}{\Delta r^3_{m1}}+\dfrac{\Delta\mathbf{r}_{m0}}{\Delta r^3_{m0}}& \hdots&  -\dfrac{\Delta\hat{\mathbf{r}}_{mn}}{\Delta r^3_{mn}}+\dfrac{\Delta\mathbf{r}_{m0}}{\Delta r^3_{m0}}\\
	\end{bmatrix},\quad \mathbf{a}_{\mathbf{S}_{M_0}}=-\mu\begin{bmatrix}
	\dfrac{\Delta\mathbf{r}_{10}}{\Delta r^3_{10}} \\
	\vdots\\
	\dfrac{\Delta\mathbf{r}_{m0}}{\Delta r^3_{m0}}\\
	\end{bmatrix},
\end{equation*}
where $\Delta\mathbf{r}_{jk}=\mathbf{r}_j-\mathbf{r}_{M_k}$ is the relative position between the $j\text{th}$ data point and $k\text{th}$ mass. Note that the vector  $\mathbf{a}_{\mathbf{S}_{M_0}}$ does not depend on any element of the mascon distribution decision variable $\mathbf{y}_{\mathbf{S}_M}$.   

\subsubsection{Mascon constraints projection}

The remaining constraints are the masses interiority within the body shape and the positiveness of $\mu_{M_0}$. In order to handle these constraints, if the loss is convex, it is possible to solve for the unconstrained problem and project the solution to feasible space. The idea is that the closest feasible point to the unconstrained minimum is the optimal solution to the constrained problem. This theory can be formally expressed as
\begin{equation}
\begin{aligned}
\mathbf{y}'&=\underset{\mathbf{y}}{\text{argmin}}\>L(\mathbf{y}),\\
\underset{\mathbf{y}\in\mathcal{Y}}{\text{argmin}}\>L(\mathbf{y})&=\underset{\mathbf{y}\in\mathcal{Y}}{\text{argmin}}\>\lVert\mathbf{y}-\mathbf{y}'\rVert_2,
\end{aligned}\label{eq:projection}
\end{equation}
where $\mathbf{y}'$ denotes the unconstrained minimum and $\mathcal{Y}$ is the feasible domain. When related to the problem under consideration, it is very likely that Eq.~\eqref{eq:loss} loss is not convex in $\mathbf{y}_{\mathbf{S}_M}$ space. However, if small steps are given towards minimum, the projection of Eq.~\eqref{eq:projection} can be applied after each step being valid since the local minimum neighborhood is possibly convex. 

Depending on the problem, Eq.~\eqref{eq:projection} projection can be as complex as directly solving the constrained optimization. Nonetheless, it is very useful for the mascon model under consideration. This is due to the fact that the remaining constraints are highly decoupled in terms of decision variables. Let consider the positiveness of the $0\textup{th}$ point mass. This mass becomes negative if the sum of the remaining $n$ masses exceeds $\mu$. It is more intuitive to visualize the situation in  space in $\sqrt{\mu_{M_k}}$ because the constraint represents a hypersphere of radius $\sqrt{\mu}$. Then, each time the mass distribution abandons that hypersphere, the projection sends the distribution back to the closest point lying on its surface. This is expressed as
\begin{equation}
\sqrt{\pmb{\mu}}_{\mathbf{S}_M}=\begin{cases}
	\sqrt{\pmb{\mu}}'_{\mathbf{S}_{M}} & \text{if}\quad \mu_{M_0}\geq0,\\
	(\sqrt{\mu}/\lVert\sqrt{\pmb{\mu}}'_{\mathbf{S}_{M}}\rVert)\sqrt{\pmb{\mu}}'_{\mathbf{S}_{M}} & \text{if}\quad\mu_{M_0}<0,\\
\end{cases}\label{eq:0thmass_constraint}
\end{equation}
which is equivalent to reduce all masses in the same proportion until $\mu_{M_0}=0$.

The interior constraint is more difficult to tackle given the complex shape of a small body. Both an interior condition and a function computing the closest point on the surface are needed. These can be derived from a small body shape polyhedron model. From \cite{Werner1996}, the polyhedron normalized Laplacian $\nabla^2U_{\text{poly}}=-\sum\limits_{f\in\text{faces}}w_f$ expression determines if a point is interior to the shape. If the point is interior the normalized Laplacian equals $-4\pi$ while it vanishes for an exterior point. The term $w_f$ is the solid angle of a polyhedron face as viewed from the evaluation point. When a violation occurs, the closest point on the polyhedron surface can be computed in a discretized form. Since the polyhedron shape is characterized by discrete surface features based on vertexes and faces, the distance of the exterior point with respect to them (or a derived variable) can be computed. Then, the closest point on the surface can be extracted. In this case, the projection condition is
\begin{equation}
	\mathbf{r}_{M_k}=\begin{cases}
		\mathbf{r}'_{M_k} & \text{if}\quad \sum\limits_{f\in\text{faces}}w_f(\mathbf{r}'_{M_k})=4\pi,\\
		\underset{\mathbf{r}_{f}}{\text{argmin}}\>\lVert\mathbf{r}_f-\mathbf{r}'_{M_k}\rVert & \text{if}\quad \sum\limits_{f\in\text{faces}}w_f(\mathbf{r}'_{M_k})=0,\\
	\end{cases}\label{eq:interior_constraint}
\end{equation}
where $\mathbf{r}_f$ refers to the center of each polyhedron face (such to only loop over the faces). Since evaluating all the polyhedron faces is computationally expensive, a low resolution polyhedron shape model may be used. 

\subsection{Mascon optimization}

A first-order technique to find the mascon distribution that minimizes Eq.~\eqref{eq:loss} is gradient descent. It also works well with the constraints projection of Eq.~\eqref{eq:0thmass_constraint}-\eqref{eq:interior_constraint} if these are checked after each update. Gradient descent requires an initial mascon distribution that is detailed below. It is also benefitial to adapt the training process, thus Adam gradient descent is tailored for this application.

\subsubsection{Initial mascon distribution}

\begin{figure}[] 
	\begin{center}
		\includegraphics[width=6cm,height=6cm,keepaspectratio,trim={1.8cm 2.7cm 1.6cm 2.5cm},clip]{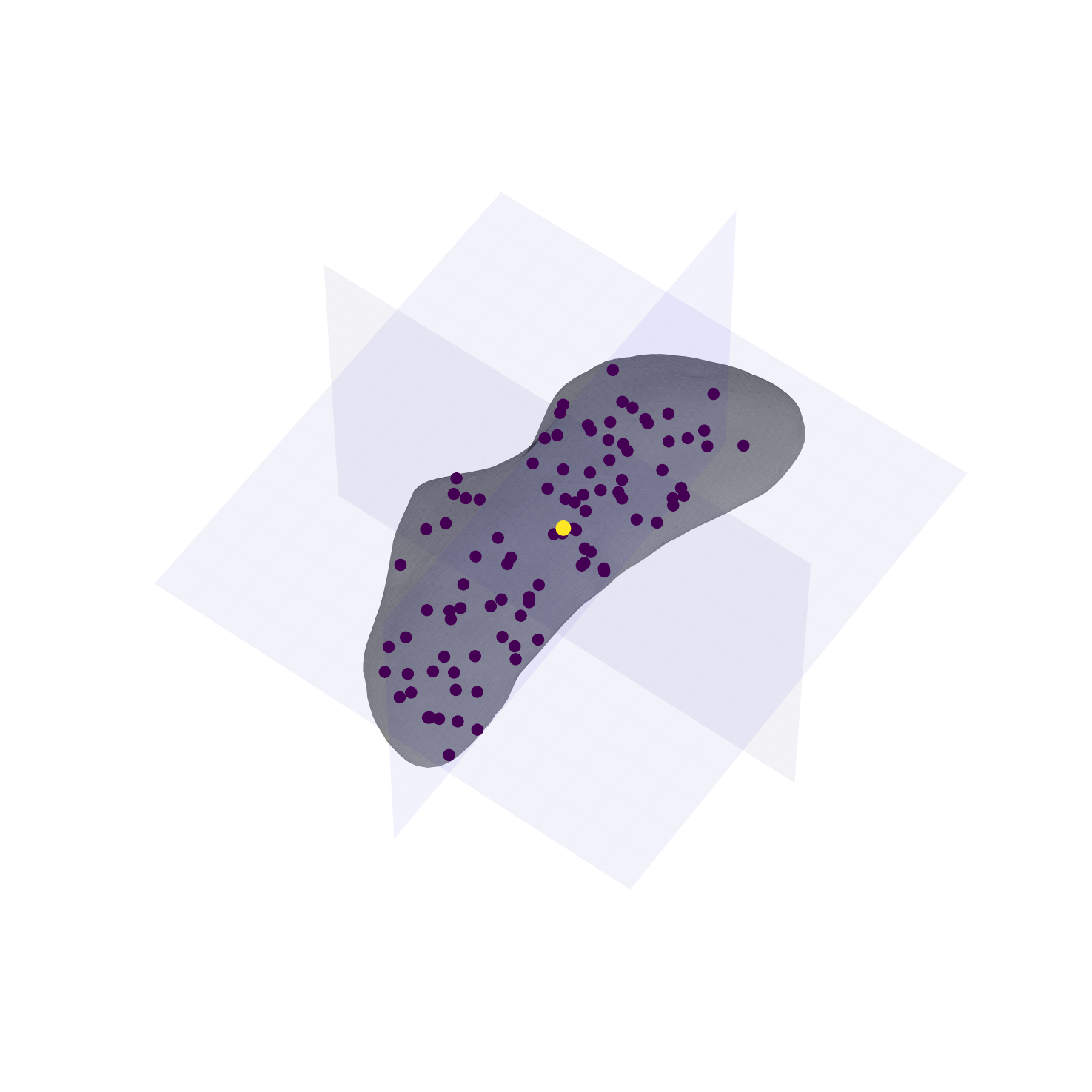}
	\end{center}
	\vspace{0cm}
	\caption{Initial mascon distribution for $n=100$.}	
	\label{fig:masconInit}
\end{figure}

In order to start the training process, gradient descent needs an initial mascon distribution. In the authors previous work \cite{Sanchez2023}, where only the mass value is fitted, the body shape is divided into eight octants. Then, an equal number of masses $(n+1)/8$ is randomly placed within each octant. This initial distribution is also used in this work (see Fig.~\ref{fig:masconInit}). Its main advantage is that the points tend to distribute well within the shape. Regarding the masses values, it is decided to start by concentrating the total mass at the origin (yellow point in Fig.~\ref{fig:masconInit}). Accordingly, the remaining $n$ masses are null. This may seem as a non-sensible choice, compared to assign $\mu/(n+1)$, but it provides consistency for trainings with different $n$. Additionally, it demonstrates how the gravity determination evolves from a basic Keplerian model.  

\subsubsection{Adam gradient descent}

The key idea behind gradient descent is to reach the local minimum by taking repeated steps opposed to the loss gradient. Still, how this basic concept is implemented has a huge impact in the solution convergence. The recent need for efficient neural network training (which is an optimization process) has led to modified versions of the classic gradient descent algorithm. One approach that has spread out within the machine-learning community is the Adam optimizer \cite{Kingma2015}. The Adam optimizer adapts the per-parameter learning rate while also adding a momentum alike term. The per-parameter adaptation is very convenient in this application because $\mu_M$ and $\mathbf{r}_M$ have a different relative impact on the acceleration prediction.  

A vital step to use gradient descent is the loss function gradient $\nabla L$ expression with respect to the decision variables. The derivation of the term $\nabla L$, for Eq.~\eqref{eq:loss} loss, is detailed in \ref{app:loss}. Then, the Adam training loop is detailed in Algorithm~\ref{alg:graddescent} (see \cite{Kingma2015} for specific details). The Adam optimizer estimates the loss gradient first $\mathbf{m}$ and second $\mathbf{v}$ moments in step 6-7. The update rule in step 8 effectively adapts the learning rate per each parameter based on the gradient moments. Note that step 8 is an element-wise division. The adaptation allows to explore parameters that may not receive updates in classic gradient descent. The momentum component appears by the gradient moving average in the update. The mascon constraints projection Eq.~\eqref{eq:0thmass_constraint}-\eqref{eq:interior_constraint} is checked after each update. If the constraints are violated, the projection step sends the decision variables to feasible space. Adam gradient descent uses the hyperparameters $\{\eta,\beta_1,\beta_2,\epsilon\}$. The variable $\eta$ is the classic learning rate. The terms $\beta_1,\beta_2\in[0,1)$ control the exponential decay rates of the gradient moments. Lastly, $\epsilon$ is a small number that ensures numerical stability in the update rule (it precludes divisions by zero).
\begin{algorithm}[]\caption{Adam optimizer with constraints projection}\label{alg:graddescent}
	\Begin{$\mathbf{y}^{[0]}_M\leftarrow\mathbf{y}_{M_0}$ (Initialize the decision variable)\;
		$\mathbf{m}^{[0]}\leftarrow\mathbf{0}$, $\mathbf{v}^{[0]}\leftarrow\mathbf{0}$ (Initialize biased first and second-order gradient moments)\;
		\For{$i\gets1$ \KwTo $i_\textup{max}$}{
			$\nabla L^{[i-1]}=\nabla L(\mathbf{y}^{[i-1]}_M)$ (Compute loss gradient)\;
			$\mathbf{m}^{[i]}\leftarrow \beta_1\mathbf{m}^{[i-1]}+(1-\beta_1)\nabla L^{[i-1]}$, $\mathbf{v}^{[i]}\leftarrow \beta_2\mathbf{v}^{[i-1]}+(1-\beta_2)(\nabla L^{[i-1]})^2$ (Update biased moments)\;
			$\hat{\mathbf{m}}^{[i]}\leftarrow \mathbf{m}^{[i]}/(1-\beta^i_1)$, $\hat{\mathbf{v}}^{[i]}\leftarrow \mathbf{v}^{[i]}/(1-\beta^i_2)$ (Correct bias)\;
			$\mathbf{y}_M^{[i]}= \mathbf{y}_M^{[i-1]}-\eta\hat{\mathbf{m}}^{[i]}/(\sqrt{\hat{\mathbf{v}}^{[i]}}+\epsilon)$ (Update decision variable)\;
			Apply Eq.~\eqref{eq:0thmass_constraint} (0th mass positiveness)\;
			\For{$k\gets1$ \KwTo $n$ $(\textup{Loop through masses})$}{
				Apply Eq.~\eqref{eq:interior_constraint} (Interior constraint)\;
			}
	}}
\end{algorithm}

It is also convenient (for the sake of numerical stability) to internally normalize some training variables. In particular, the masses are divided by $\mu/(n+1)$. To normalize the position, a tenth of the body shape elongation on each direction is used. Note that each position coordinate is scaled differently (which helps to treat elongated small bodies). Finally, the gravity accelerations are normalized by the norm of the $j\text{th}$ data point as $\mathbf{a}_M/\lVert \mathbf{a}_j\rVert_2$ and $\mathbf{a}_j/\lVert \mathbf{a}_j\rVert_2$ respectively.   

\subsection{Simultaneous navigation and gravity estimation scheme}

\begin{figure}[] 
	\begin{center}
		\includegraphics[width=11cm,height=11cm,keepaspectratio,,trim={0 0cm 0 0cm},clip]{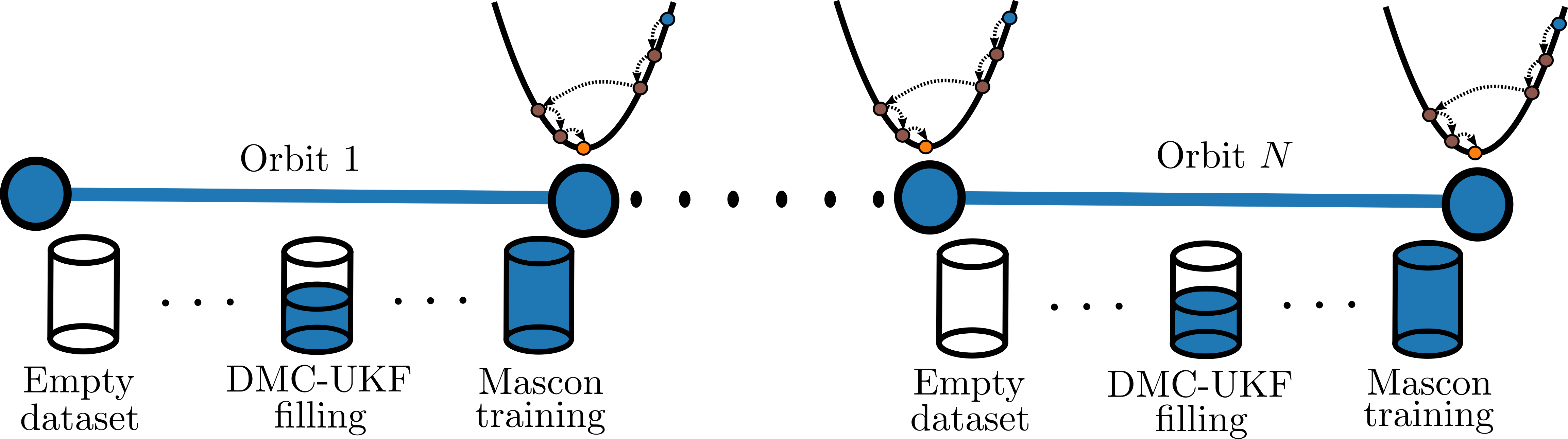}
	\end{center}
	\vspace{-0.5cm}
	\caption{Illustration of the gravity estimation process.}	
	\label{fig:BSK_gravprocess}
\end{figure}

The simultaneous navigation and mascon gravity estimation scheme links the previous paragrahps with Section~\ref{sec:DMC-UKF} DMC-UKF. Recall that the first step is to generate a position-unmodeled acceleration dataset using Algorithm~\ref{alg:UKF_pseudocode} DMC-UKF. Then, the dataset is used to fit a mascon distribution using Adam Algorithm~\ref{alg:graddescent}. Still, how to manage this strategy over time is to be chosen. In this work, it is chosen to process the data in 1-orbit batches. This means that the DMC-UKF fills the dataset during one orbit. At the end of this orbit, the mascon distribution is fitted with the previous data. Subsequently, the trained mascon distribution is uploaded to the DMC-UKF and the data batch is emptied. This process is repeated until the final orbit as shown in Fig.~\ref{fig:BSK_gravprocess}.

\begin{figure}[] 
	\begin{center}
		\includegraphics[width=12cm,height=12cm,keepaspectratio]{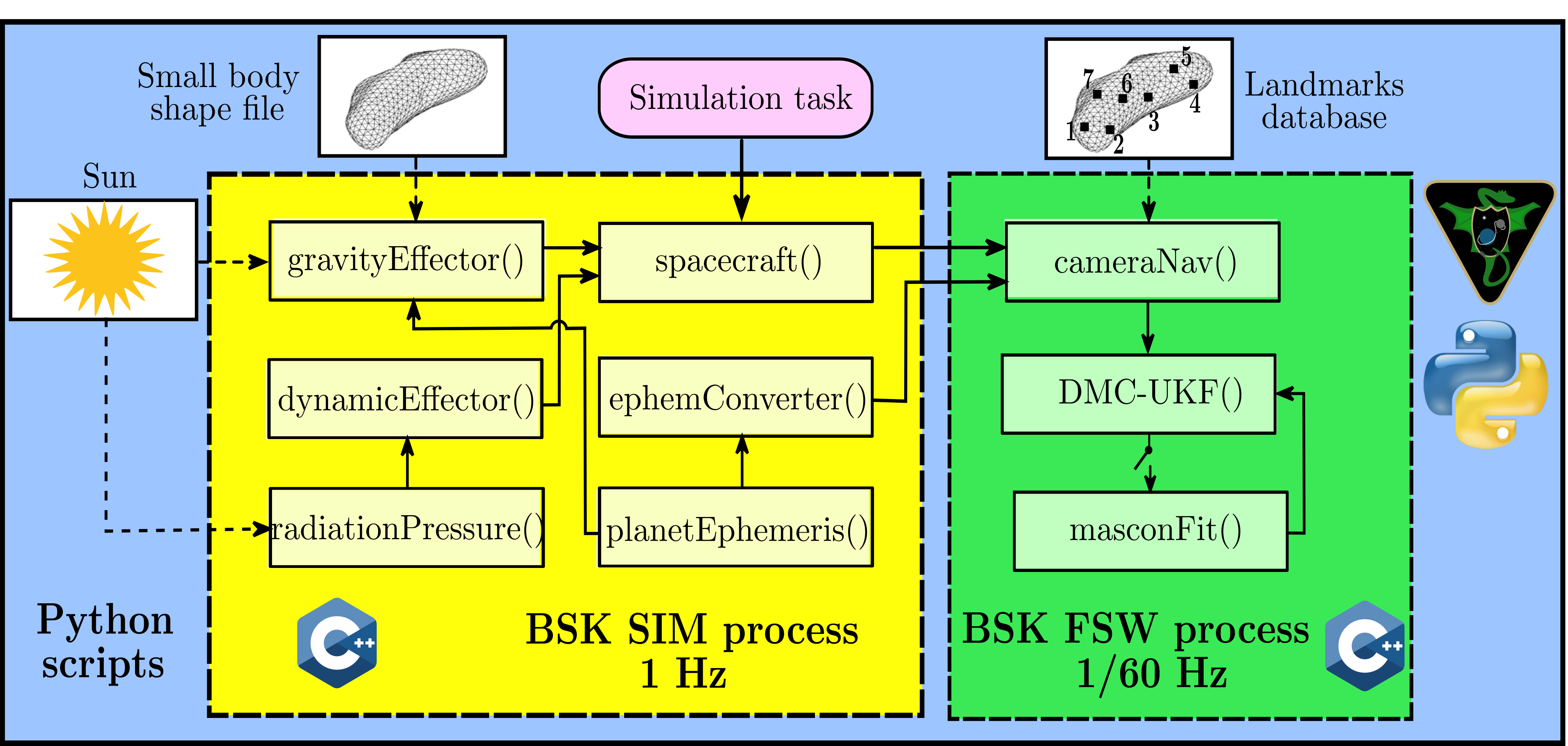}
	\end{center}
	\vspace{-0.5cm}
	\caption{Diagram of simultaneous navigation and gravity estimation in the Basilisk simulation framework.}	
	\label{fig:BSK_diagram}
\end{figure}

The strategy is implemented using the  Basilisk\footnote[1]{\url{http://hanspeterschaub.info/basilisk/}} (BSK) software. BSK is an open source astrodyn cacta amics simulation tool \cite{Kenneally2022} composed of Python modules coded in C/C++. This offers the convenience of managing the simulation with Python scripts while benefiting of the background C/C++ execution speed. The BSK implementation of the simultaneous navigation and mascon gravity estimation is illustrated in Fig.~\ref{fig:BSK_diagram}. It consists of a ground truth simulator and a flight software process. The ground truth simulator uses a fixed time step 4th order Runge-Kutta method to integrate Eq.~\ref{eq:spacecraft_dynamics} spacecraft dynamics. The flight software process encompasses the landmarks pixel generator (cameraNav), the DMC-UKF and the mascon optimizer. The pixel generator and DMC-UKF are processed at the same frequency while the mascon optimizer is executed after each orbit. 

\section{Results}\label{sec:results}

This section tests the simultaneous navigation and mascon gravity estimation scheme through numerical simulations. The results are divided into three subsections. In the first place, only the mascon optimizer with true data. This validates the fitting of the mascon distribution. Subsequently, the simultaneous navigation and gravity estimation results are presented. Finally, a propagation analysis is done with the results of previous subsections. 

Due to the extensive data available from NEAR mission, the asteroid 433 Eros is chosen as the target small body. Eros has a standard gravity parameter $\mu=4.4627547\cdot10^5~\text{m}^3/\text{s}^2$ and a rotational period $T_A=5.27~\text{h}$. Its ground truth gravity field is modeled using a polyhedron shape with 7790 triangular faces\footnote[2]{\url{https://sbnarchive.psi.edu/pds3/near/NEAR_A_5_COLLECTED_MODELS_V1_0/data/msi/}}. Eros heliocentric orbital parameters are $\{a=1.4583~\text{AU},e=0.2227,i=10.829^{\circ}, \Omega=304.4^{\circ}, \omega=178.9^{\circ},\nu_0=246.9^{\circ}\}$ and its orientation is defined by $\{\text{RA}=11.369^{\circ},\text{dec}=17.227^{\circ},\text{LST}_0=0^{\circ}\}$. The orbiting spacecraft is physically characterized by a mass $m=750~\text{kg}$, a reflection coefficient $C_R=1.2$ and a solar radiation pressure exposed area $S=1.1~\text{m}^2$. 

\subsection{On-orbit data impact on gravity estimation}\label{sec:onorbitruth}

This paragraph analyzes the mascon gravity estimation algorithm of Section~\ref{sec:gravity}. The differences between a dense and an on-orbit dataset are emphasized. Using the previous data, the global gravity accuracy of a complete mascon training and a static one (just fit the masses) are stated.   

Recall that the Adam optimizer (see Algorithm~\ref{alg:graddescent}) requires choosing four hyperparameters and the number of iterations. The hyperparameters are tuned ad hoc as $\{\eta=10^{-3},\beta_1=0.9,\beta_2=0.99,\epsilon=10^{-6}\}$ while 1000 iterations are used. These values are used throughout the rest of the manuscript.

\subsubsection{Training datasets}

The dataset plays a fundamental role in the mascon gravity accuracy. Just using on-orbit data is challenging because the relevant low altitude features are missed. This may distort the maximum potential accuracy of the mascon gravity model. In order to obtain a bigger picture, a dense dataset homogeneously covering the small body is also considered. Furthermore, the accuracy gap between the dense and on-orbit dataset can be shrunk by the inclusion of low altitude data. This could be associated to ejecta (or gravity trackers \cite{Villa2021}) measurements though these are not simulated realistically. 

The on-orbit dataset is based on a spacecraft with initial orbital elements as $\{a_0=34~\text{km},e_0=0.001, i_0=45^{\circ}, \Omega_0=48.2^{\circ}, \omega_0=347.8^{\circ},\nu_0=85.3^{\circ}\}$. This dataset is composed of position-acceleration components sampled each $60~\text{s}$ along 10 orbital periods  ($\sim 982$ data points per orbit). Since the idea is to mimic conditions to be found in the subsequent paragraph \ref{sec:results:simultaneous}, two considerations are made. The first one is that the gravity estimation process follows the sequential approach of Fig.~\ref{fig:BSK_gravprocess}. This means that each orbit batch is fitted sequentially by using the solution of the previous orbit as the starting point for the next one. The second consideration is the removal of points where no landmarks are visible due to lighting conditions ($\approx20\%$ reduction of the samples). Alternatively, the dense dataset is generated by randomly placing data (same number of samples as the on-orbit) between the surface and up to $30~\text{km}$ radius. The training with the dense dataset also follows the sequential batch processing. 

\begin{figure}[htbp] 
	\begin{center}
		\includegraphics[width=12cm,height=12cm,keepaspectratio,trim={0cm 0.15cm 0 0.1cm},clip]{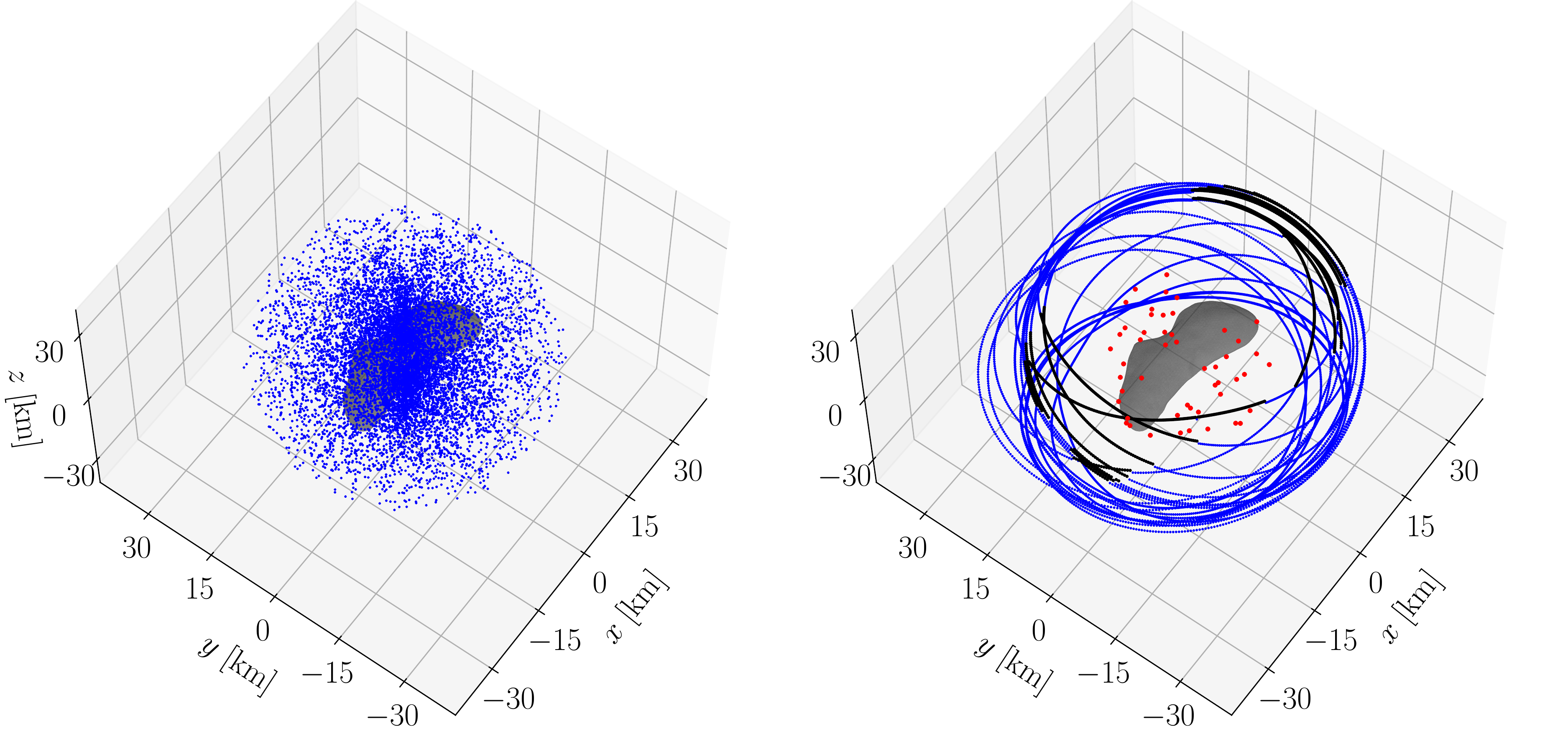}
	\end{center}
	\vspace{0cm}
	\caption{Dense ($\textit{left}$) and on-orbit ($\textit{right}$) datasets. Black $\equiv$ points removed; red $\equiv$ low altitude samples.}	
	\label{fig:datasets}
\end{figure}

The dense and on-orbit datasets are depicted in Fig.~\ref{fig:datasets}. The aforementioned low altitude samples that may complement the on-orbit dataset are also shown. The low altitude data is generated randomly but its spatial diversity is limited. In particular, the low altitude samples are related to the spacecraft first orbit arc. A sample is randomly drawn for an orbit point which defines its sample latitude and longitude. Then, a random radius ranging from the surface up to $18~\text{km}$ is assigned. When considering this additional data, 50 samples ($\approx6\%$ of an orbit batch) are assumed available since the beginning of the estimation process. 

\subsubsection{Global gravity accuracy}\label{sec:orbittruth_accuracy}

\begin{figure}[htbp] 
	\begin{center}
		\includegraphics[width=8cm,height=8cm,keepaspectratio,trim={0cm 0.25cm 0 0.25cm},clip]{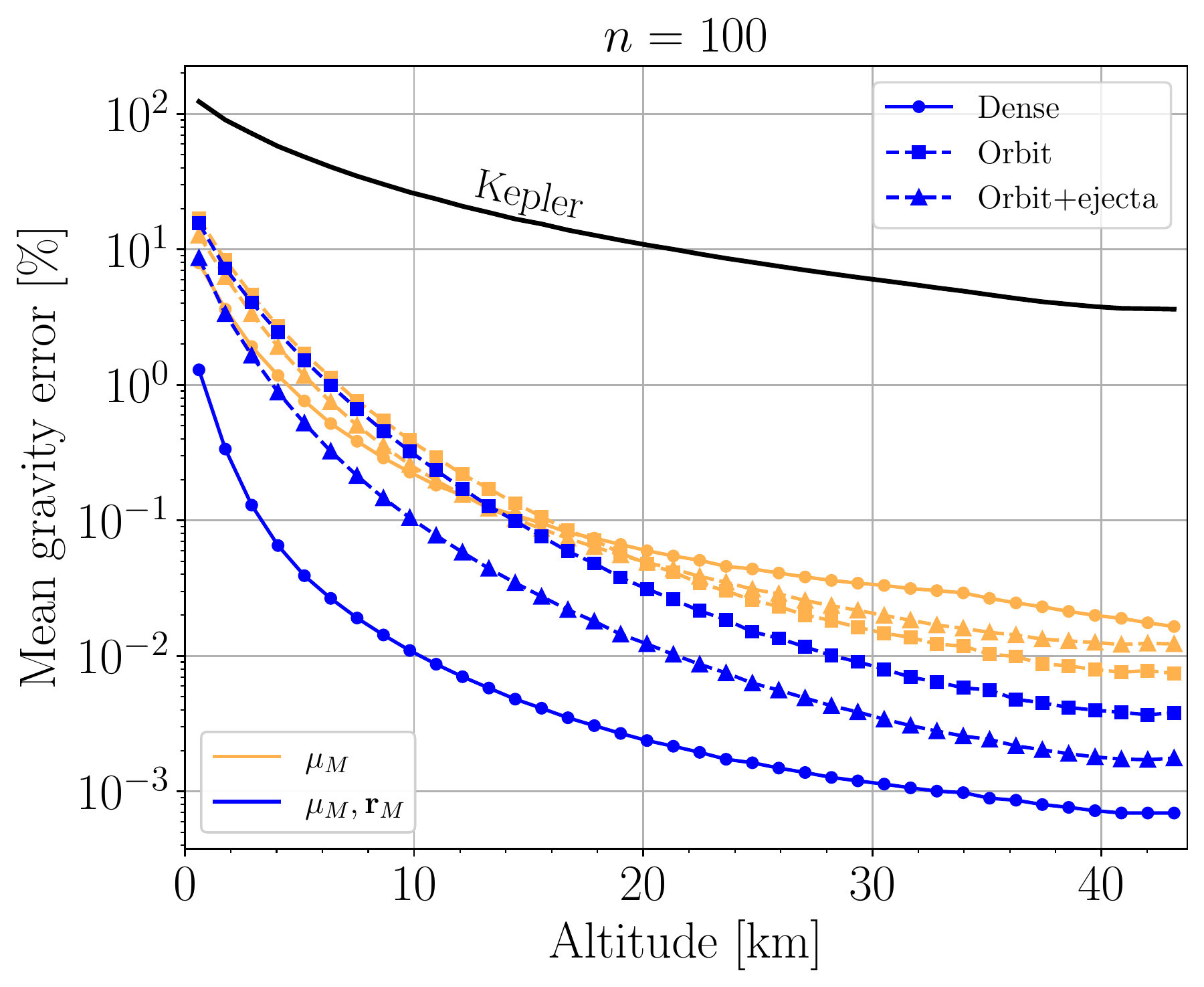}
	\end{center}
	\vspace{0cm}
	\caption{Mean gravity error with respect to altitude for dense and on-orbit datasets.}	
	\label{fig:offline}
\end{figure}

\begin{figure}[htbp] 
	\begin{center}
		\includegraphics[width=11cm,height=11cm,keepaspectratio,trim={0cm 0.25cm 0 0.25cm},clip]{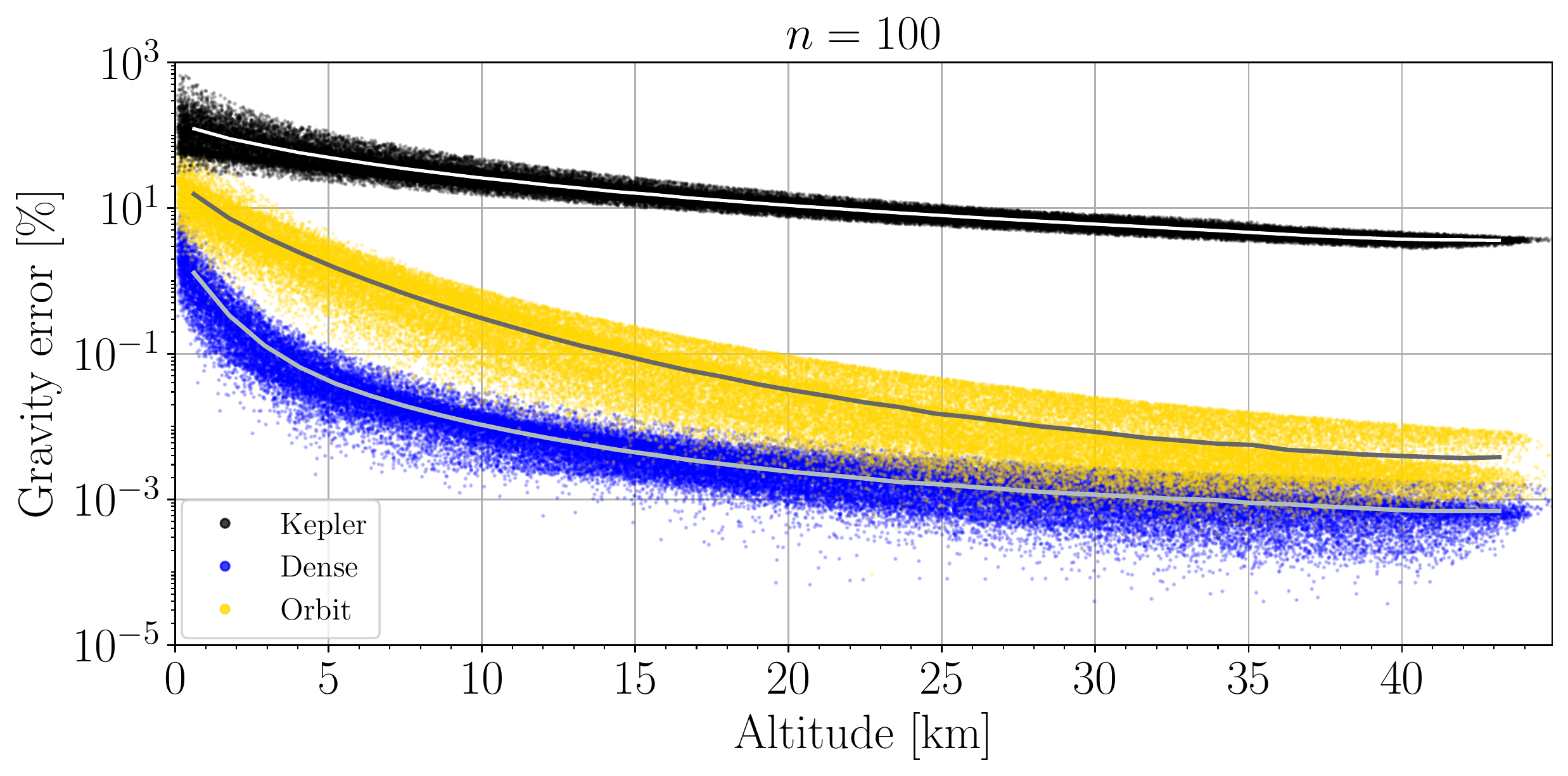}
	\end{center}
	\vspace{0cm}
	\caption{Gravity error with respect to altitude for dense and on-orbit datasets.}	
	\label{fig:offlineErrAlt}
\end{figure}

It is of interest to evaluate the fitted mascon distributions, using Fig.~\ref{fig:datasets} datasets, gravity accuracy. For this purpose, 55166 evaluation points are placed between the small body surface up to a radius of $50~\text{km}$. These evaluation points are uniformly distributed within altitude bands of $1.2~\text{km}$ (containing each $\sim1400$ points). Then, the mean gravity percent errors (see Eq.~\ref{eq:grav_error}) for these altitude bands are used as a global gravity accuracy metric. An example of this metric is shown in Fig.~\ref{fig:offline}. In order to create that figure, a mascon distribution with $n=100$ masses has been trained under different datasets (dense, on-orbit and on-orbit with low altitude samples). The low altitude samples are referred as ejecta in the legends. Furthermore, static mascon distributions (just fitting $\sqrt{\mu_M}$) are also trained. The error with a simple Keplerian gravity is included as a reference. Several conclusions can be derived from Fig.~\ref{fig:offline}. The mascon training under the dense dataset is highly accurate with a maximum mean gravity error of $\approx1\%$ very close to the surface. Instead, the use of the on-orbit dataset increase the altitude bands error in an order of magnitude with respect to the dense dataset. The inclusion of the low altitude samples has a positive effect in driving down the on-orbit errors. Lastly, the trainings using static mascon distributions show to be less accurate than the complete ones. For the static distributions, the inclusion of low altitude samples increase accuracy in that domain but degrades the high altitudes. When these low altitude vanishes (on-orbit), the complete and static mascon distribution achieve similar accuracy in the low altitude domain. This also highlights that the complete mascon training requires more information to acquire its maximum potential. Each line in Fig.~\ref{fig:offline} compresses considerable information. To illustrate what these lines are representing, Fig.~\ref{fig:offlineErrAlt} plots the gravity error with respect to each evaluation point altitude. For the sake of clarity, this is only done for the complete mascon distribution trainings with the dense and on-orbit datasets (and the Kepler model).

\begin{figure}[htbp] 
	\begin{center}
		\includegraphics[width=8cm,height=8cm,keepaspectratio,trim={0cm 0.25cm 0 0.25cm},clip]{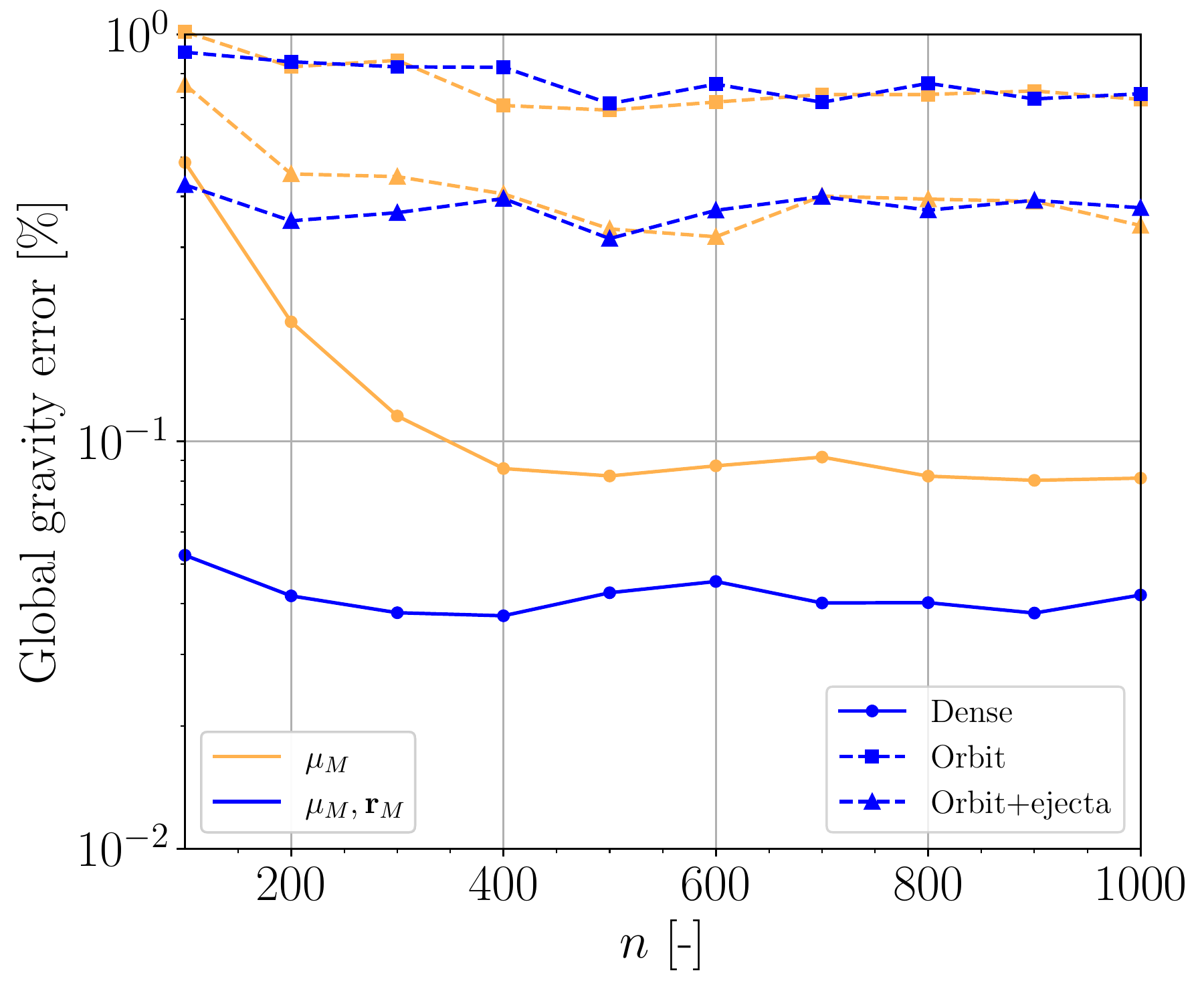}
	\end{center}
	\vspace{0cm}
	\caption{Global gravity error with respect to number of masses for different mascon setups and training datasets.}	
	\label{fig:offlinenM}
\end{figure}

Another question that emerges is the effect of the number of masses $n$ on the gravity accuracy. To answer this query, $n$ is varied between 100 and 1000 under the previous training conditions. Since a new variable ($n$) is added to the analysis, the gravity error is compressed to the global one (mean gravity for the entire evaluation set). The result is shown in Fig.~\ref{fig:offlinenM}. It can be observed that the training varying both the mascon distribution position and mass is relatively indifferent to $n$. On the contrary, $n$ seems to be relevant (up to $n=400$) when only the mascon masses are trained. This occurs for both the dense dataset and the on-orbit one with low altitude samples. Additionally, training the mascon distribution positions does not seem to grant a distinctive advantage for the pure on-orbit dataset.

\subsection{Simultaneous navigation and gravity estimation}\label{sec:results:simultaneous}

Next the results with the simultaneous navigation and mascon gravity estimation scheme (see Fig.~\ref{fig:BSK_diagram}) are presented. The simulated orbit corresponds to the one shown in Fig.~\ref{fig:datasets} (\textit{right}). In this case, several parameters for the DMC-UKF are to be provided. 

The DMC-UKF tuning parameters follow the ones used in \cite{Wan2000} which are $\{\alpha=0,\beta=2,\lambda=10^{-3}\}$. The DMC-UKF is initialized with true initial position and velocity while the unmodeled acceleration is assumed null. Moreover, the unmodeled acceleration is reinitialized to a null value after a gravity fit is completed or a measurement outage arises. The initial state covariance and process noise are 
\begin{equation}
	\begin{aligned}
		\pmb{\Sigma}_{\mathbf{x}\mathbf{x}}(t_0)&=\begin{bmatrix}
			10^2\mathbf{I}\>\>(\text{m})^2 & \mathbf{0}_{3\times3} & \mathbf{0}_{3\times3}\\
			\mathbf{0}_{3\times3} & 10^{2}\mathbf{I}\>\>(\text{mm/s})^2  & \mathbf{0}_{3\times3}\\
			\mathbf{0}_{3\times3} & \mathbf{0}_{3\times3} & 1^{2}\mathbf{I}\>\>(\mu\text{m/s}^2)^2 \\
		\end{bmatrix},\\
		\pmb{\Sigma}_{\mathbf{f}\mathbf{f}}&=\begin{bmatrix}
			0.1^2\mathbf{I}\>\>(\text{m})^2 & \mathbf{0}_{3\times3} & \mathbf{0}_{3\times3}\\
			\mathbf{0}_{3\times3} & 1^{2}\mathbf{I}\>\>(\text{mm/s})^2  & \mathbf{0}_{3\times3}\\
			\mathbf{0}_{3\times3} & \mathbf{0}_{3\times3} & 2^{2}\mathbf{I}\>\>(\mu\text{m/s}^2)^2 \\
		\end{bmatrix}.
	\end{aligned}
\end{equation}  
In Kalman filters, adequate tuning of the process noise is key for accuracy. In this case, $\pmb{\Sigma}_{\mathbf{f}\mathbf{f}}$ is determined ad hoc via experimentation. It is noticed that the process uncertainty associated to $\mathbf{a}$ is the parameter that requires a finer tuning.   

Regarding measurements, navigation data of 100 surveyed landmarks is assumed available. These are placed at the face centers of Eros polyhedron model. The landmarks are tracked by the on-board camera with 4:3 aspect ratio, a sensor size of $17.3\times13~\text{mm}$ and a resolution of $2048\times1536~\text{px}$. Consequently, the pixel width is $w_p=8.447~\mu\text{m}$. The camera focal length is $f=25~\text{mm}$ which is derived by a static position determination analysis in \ref{app:camera}. Landmark measurements are sampled each $60~\text{s}$. The DMC-UKF uncertainty on each measurement pixel is tuned to be the unity, thus
\begin{equation}
\pmb{\Sigma}_{\mathbf{z}\mathbf{z}_{j}}=\begin{bmatrix}
1^2\>\>(\text{px})^2& 0\\
0 & 1^2\>\>(\text{px})^2\\
\end{bmatrix},\quad \pmb{\Sigma}_{\mathbf{z}\mathbf{z}}=\begin{bmatrix}
\pmb{\Sigma}_{\mathbf{z}\mathbf{z}_{1}} & \hdots & \mathbf{0}_{2\times2}\\
\vdots & \ddots & \vdots\\
 \mathbf{0}_{2\times2} & \hdots & \pmb{\Sigma}_{\mathbf{z}\mathbf{z}_{n_L}}\\
\end{bmatrix},
\end{equation}
where the measurement noise matrix $\pmb{\Sigma}_{\mathbf{z}\mathbf{z}}$ dimension changes according to the number of visible landmarks $n_L$.

\subsubsection{DMC-UKF estimation}

Let assess the accuracy of the DMC-UKF estimates which compose the subsequent mascon training dataset. The results of this paragraph concerns a complete mascon distribution fitting with $n=100$. The DMC-UKF accuracy greatly depends on the simulation conditions. To test this, four different scenarios (namely A1, A2, B1 and B2) are considered. Scenarios A do not account for the lighting constraint (thus no measurements outages arise) while B ones do. The numeric label ``1'' refers to a simulation without landmarks uncertainty. Alternatively, the ``2'' means that the filter landmark database is perturbed with a $5~\text{m}$ standard deviation Gaussian error. These conditions are summarized in Table~\ref{tab:scenarioDMCUKFresults}. Additionally, the position and unmodeled acceleration root mean square errors (RMSE) are reported. This gives an idea on the severity of the different scenario conditions. As expected, scenario A1 is the most favorable one as the filter has continuous measurements and knows the exact landmark locations. The most severe conditions arise when measurement outages happen (scenarios B). This is mainly because the unmodeled acceleration has to converge again after each of these gaps.    
\begin{table}[]  
	\centering
	\begin{tabular}{|c|cc|cc|}
		\hline
		Scen. &  Lighting & Landmark error & $\text{RMSE}(r)~[\text{m}]$ & $\text{RMSE} (a)~[\%]$\\
		\hline
		A1 & Full & No & 2.164 & 1.202\\
		\hline
		A2 & Full & Yes & 3.765 & 1.367\\
		\hline  
		B1 & Partial & No & 5.320 & 1.769\\
		\hline  
		B2 & Partial & Yes & 8.448 & 2.042\\          
		\hline
	\end{tabular}
	\caption{Summary of scenario conditions and position-unmodeled acceleration RMSE.}
	\label{tab:scenarioDMCUKFresults}
\end{table}

\begin{figure}[] 
	\begin{center}
		\includegraphics[width=12cm,height=12cm,keepaspectratio,trim={0 1cm 0 0.45cm},clip]{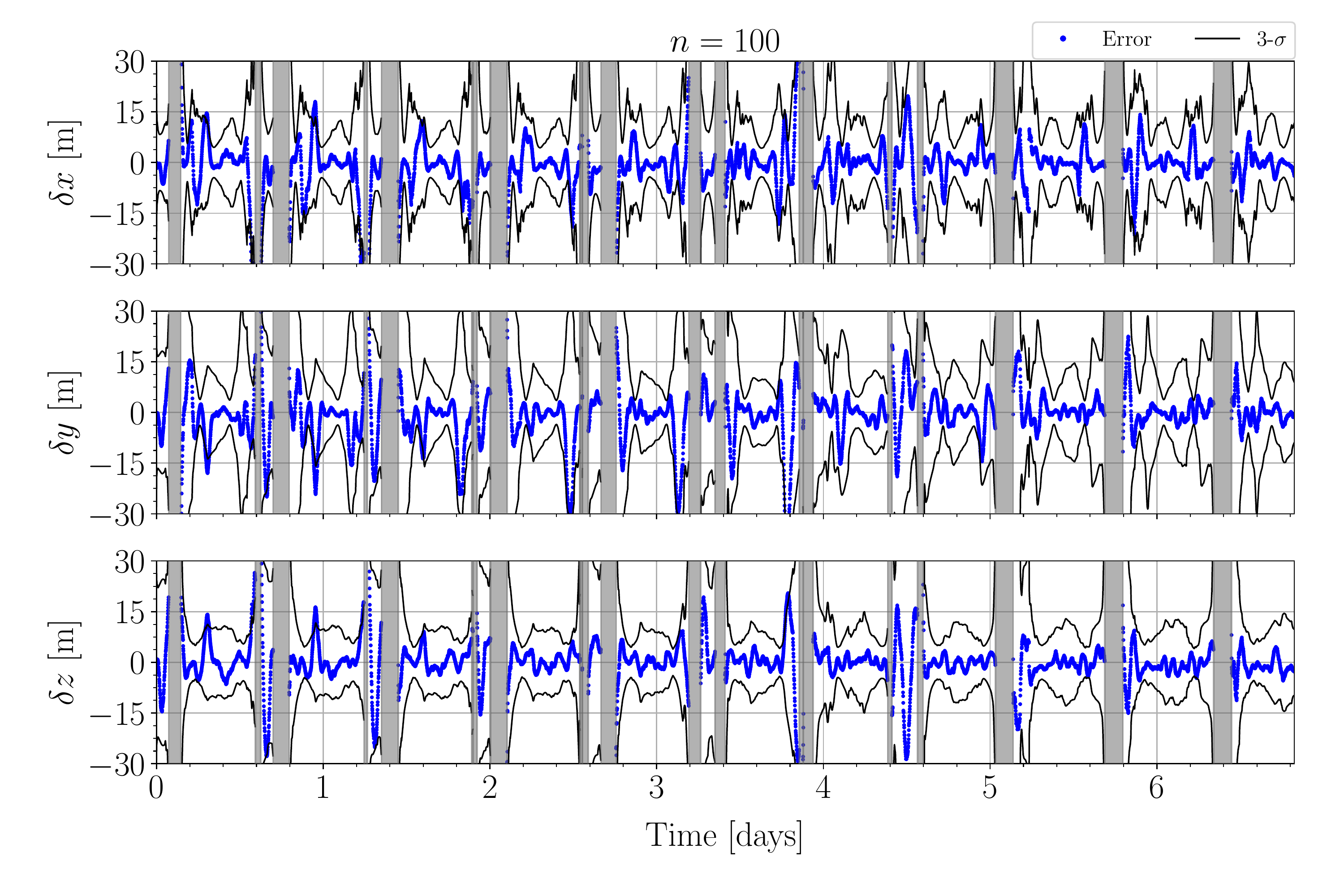}
	\end{center}
	\vspace{-0.5cm}
	\caption{DMC-UKF position error and uncertainty estimation for scenario B2. Gray area $\equiv$ measurement outage.}	
	\label{fig:poserror_DMCUKF}
\end{figure}

\begin{figure}[] 
	\begin{center}
		\includegraphics[width=12cm,height=12cm,keepaspectratio,trim={0 1.25cm 0 0.45cm},clip]{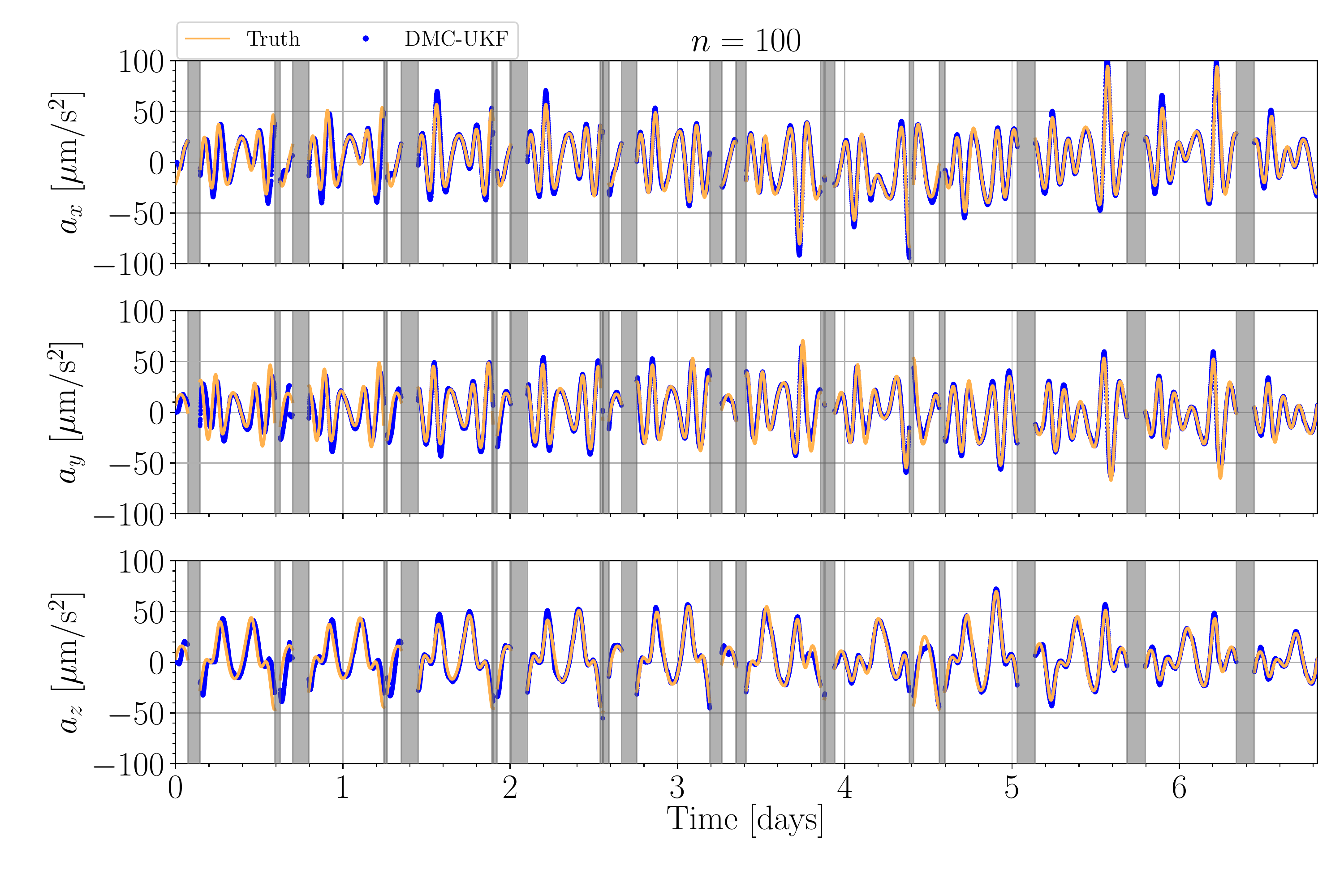}
	\end{center}
	\vspace{-0.5cm}
	\caption{DMC-UKF inhomogeneous gravity estimation for scenario B2. Gray area $\equiv$ measurement outage.}	
	\label{fig:accerror_DMCUKF}
\end{figure}

In order to illustrate the DMC-UKF signal, scenario B2 position error and inhomogeneous gravity estimate are shown in Fig.~\ref{fig:poserror_DMCUKF}-\ref{fig:accerror_DMCUKF}. The inhomogeneous gravity component is most representative as it avoids the dominant Keplerian term. In Fig.~\ref{fig:poserror_DMCUKF}, it can be observed that the loss and recapture (gray areas) of landmark measurements produce sharp errors. The position uncertainty seems to be bounded and the filter adapts itself to changes on it. The DMC-UKF is capable to keep track of the inhomogeneous gravity signal (which is the primary goal) as shown in Fig.~\ref{fig:accerror_DMCUKF}. The plotted estimate is the addition of the unmodeled acceleration estimate $\hat{\mathbf{a}}$ plus the inhomogeneous gravity component of the current filter model, $\mathbf{a}_M(\hat{\mathbf{r}})+\mu\hat{\mathbf{r}}/\hat{r}^3$. It can be observed that the DMC-UKF signal matches more closely the truth one after a couple of mascon trainings. Furthermore, spikes (due to orbital perturbations) arising in the second half of the simulation are captured to a certain extent.

\subsubsection{On-orbit global gravity accuracy}

\begin{figure}[htbp] 
	\begin{center}
		\includegraphics[width=8cm,height=8cm,keepaspectratio,trim={0cm 0.25cm 0 0.25cm},clip]{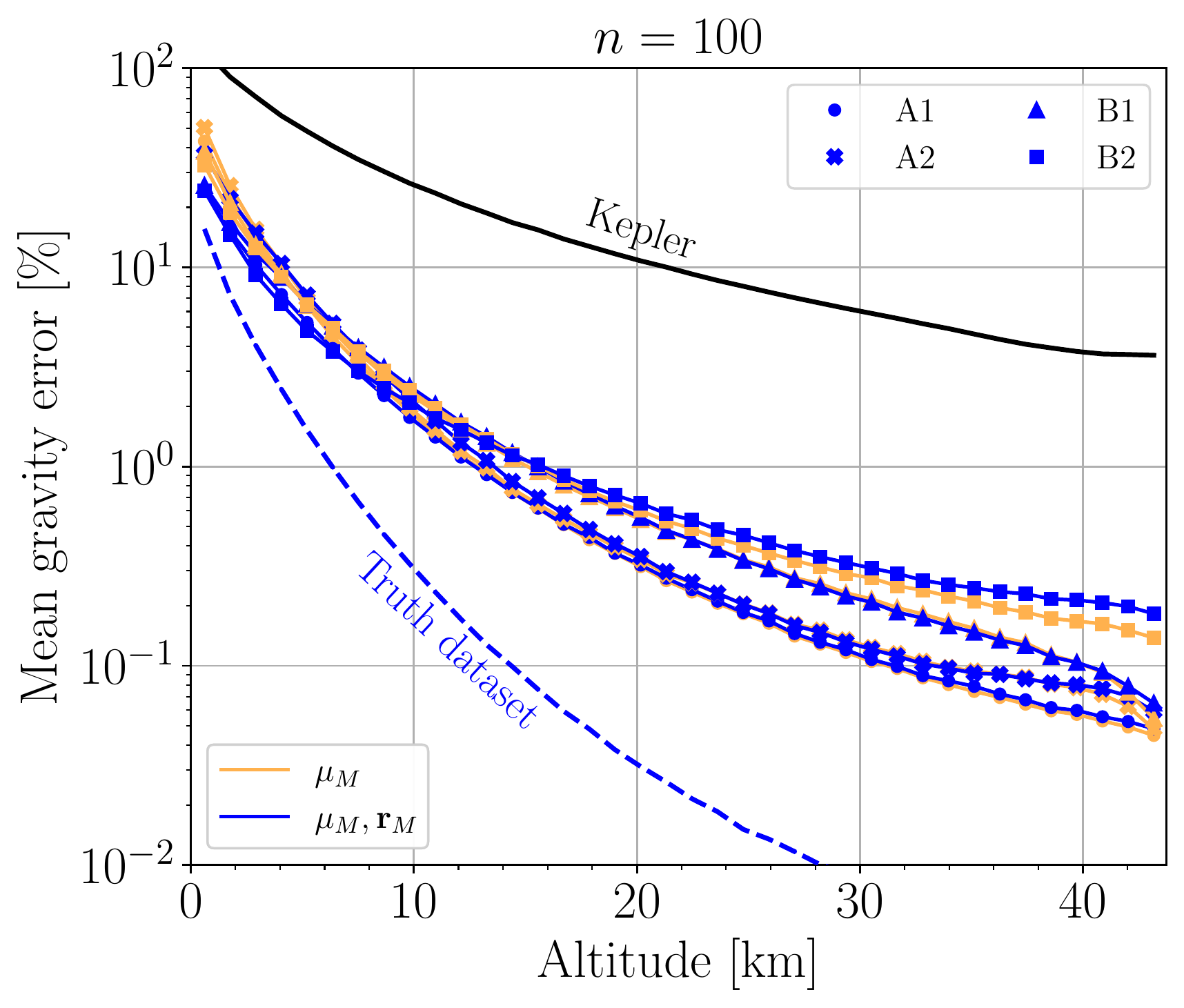}
	\end{center}
	\vspace{0cm}
	\caption{Mean gravity error with respect to altitude with simultaneous navigation and gravity estimation.}	
	\label{fig:online}
\end{figure}

\begin{figure}[htbp] 
	\begin{center}
		\includegraphics[width=8cm,height=8cm,keepaspectratio,trim={0cm 0.25cm 0 0.25cm},clip]{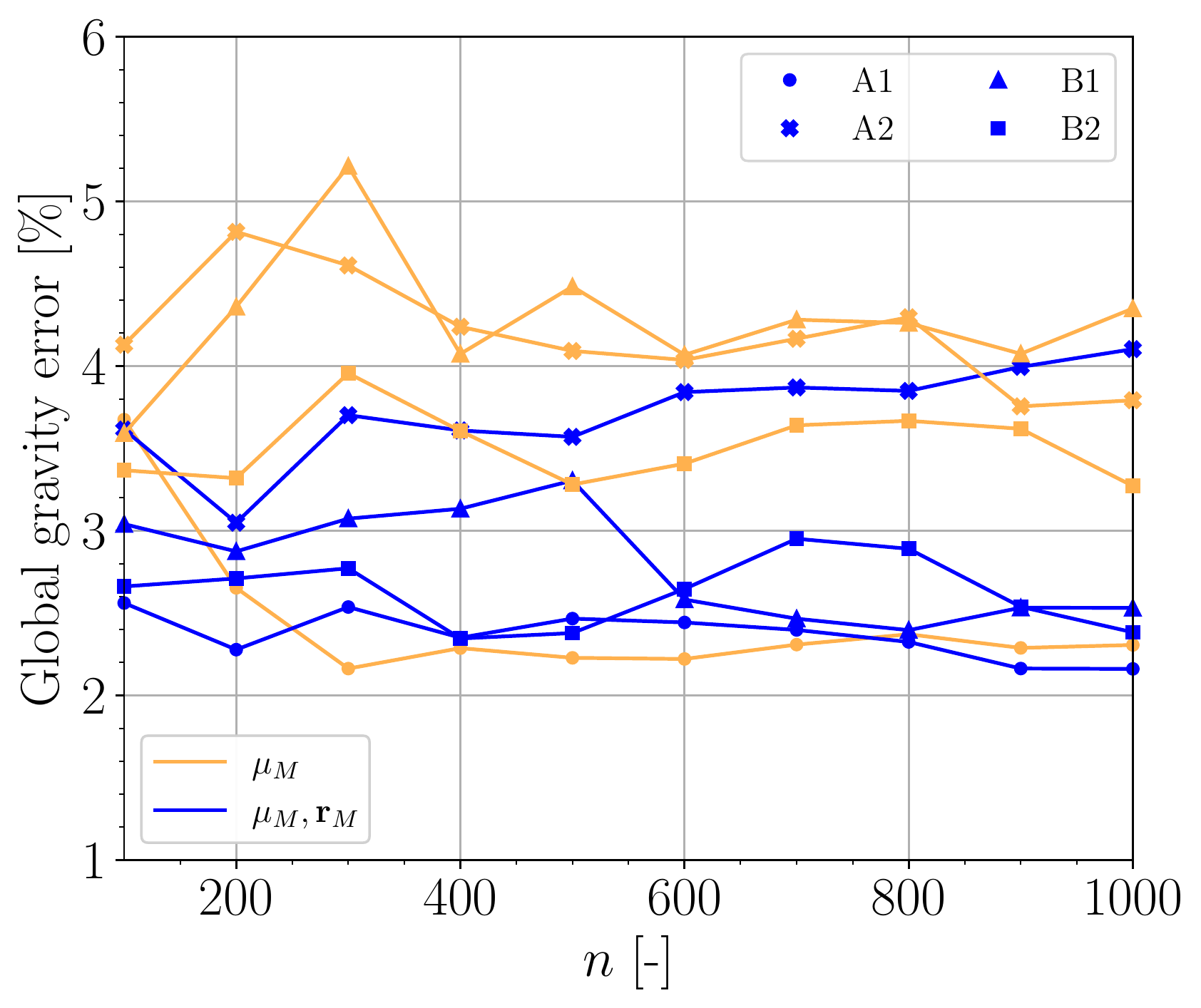}
	\end{center}
	\vspace{0cm}
	\caption{Mean gravity error with respect to altitude for different mascon setups and training datasets.}	
	\label{fig:onlinenM}
\end{figure}

The on-orbit gravity accuracy for Table~\ref{tab:scenarioDMCUKFresults} scenarios is now assessed. The evaluation set is the same as in paragraph~\ref{sec:orbittruth_accuracy}. For $n=100$, the mean gravity error with respect to altitude is shown in Fig.~\ref{fig:online} along with the Kepler model and the orbit fit with a true dataset (for reference purposes). The on-orbit training seems effective as the initial Keplerian model is improved. Nonetheless, significant degradation arises with respect to the true on-orbit dataset. The errors tends to correlate with the severity of the scenarios in Table~\ref{tab:scenarioDMCUKFresults}. In Fig.~\ref{fig:online}, the complete or the static mascon distribution does not seem to significantly differ from each other. It can be noticed that the complete training seems to perform better at low altitudes while the static one fits slightly better the high altitude domain. The analysis is complemented with Fig.~\ref{fig:onlinenM} which shows the global error with respect to the number of masses. Except from the static mascon training under scenario A1, no clear correlations are derived in terms of global error with respect to $n$. It also shows that in terms of global error the complete mascon trainings seems to be better, overall, than the static ones. However, awareness should be raised that the global gravity error favors low errors in the low altitude domain (since low altitude errors are higher in absolute terms). This is why scenario A2 (see low altitude domain error in Fig.~\ref{fig:online}), which does not have the most severe conditions, looks disfavorable in Fig.~\ref{fig:onlinenM}. 

\begin{figure}[htbp] 
	\begin{center}
		\includegraphics[width=13.5cm,height=13.5cm,keepaspectratio,trim={1.5cm 1.0cm 0 3.5cm},clip]{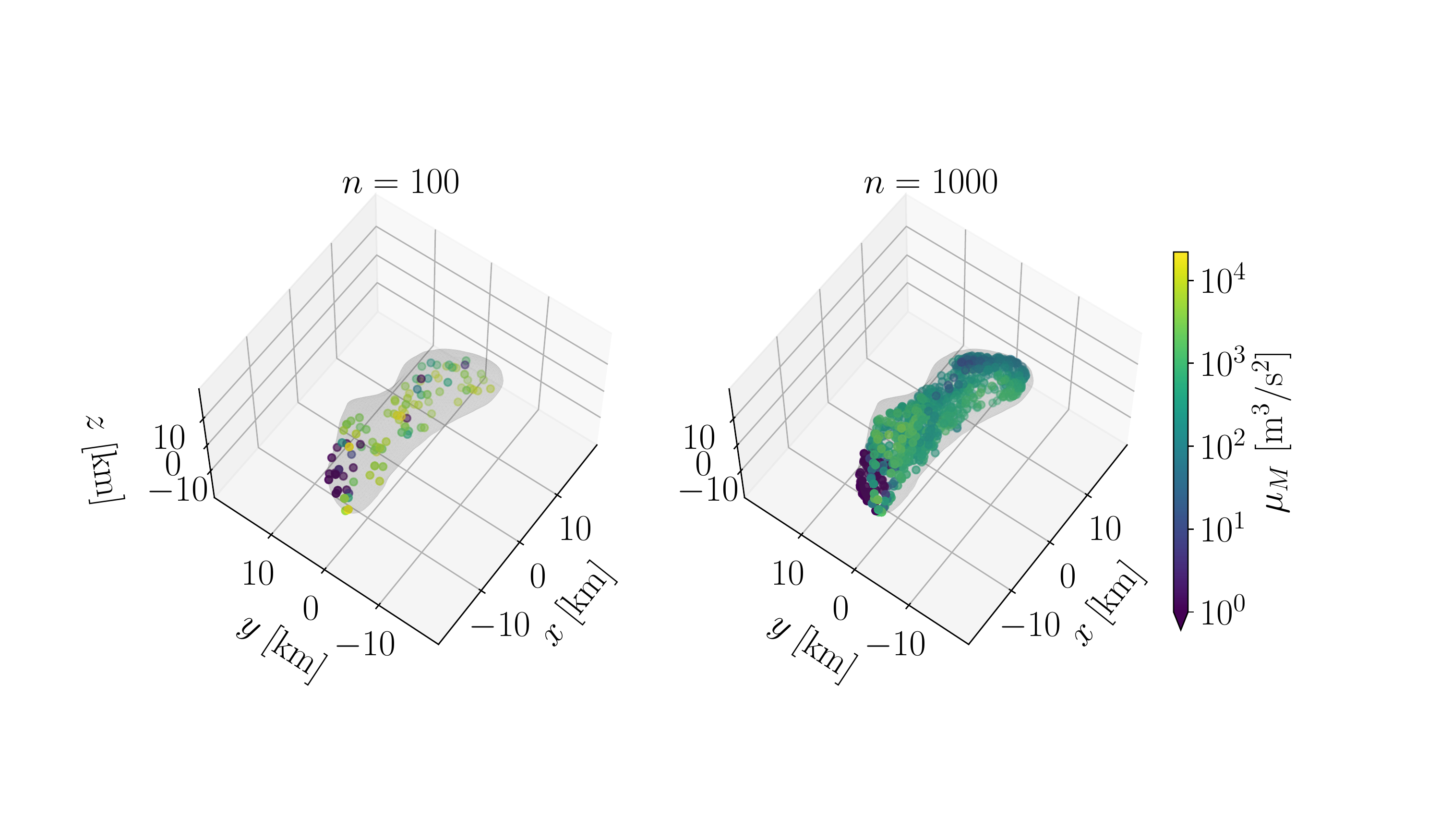}
	\end{center}
	\vspace{0cm}
	\caption{Trained mascon distribution for scenario B2.}	
	\label{fig:mascon}
\end{figure}

It may be also worth to look the trained mascon distributions. Figure~\ref{fig:mascon} shows the resulting distributions of trainings for $n=100,1000$ under scenario B2. The average gravity parameters of these distributions are $4418$ and $445.8~\text{m}^3/\text{s}^2$ respectively. It is found that some masses tend to negligible values ($<1~\text{m}^3/\text{s}^2$), especially near the most negative $x$ axis, constituting a 14.8\% ($n=100$) and a 6.89\% ($n=1000$) of the total number. Along the training, the masses positions vary in average $2.307$ and $2.741~\text{km}$ with respect to the initial distribution.

\subsubsection{Inclusion of low altitude samples}

\begin{figure}[htbp] 
	\begin{center}
		\includegraphics[width=14cm,height=14cm,keepaspectratio,trim={1.5cm 0.05cm 0 0.25cm},clip]{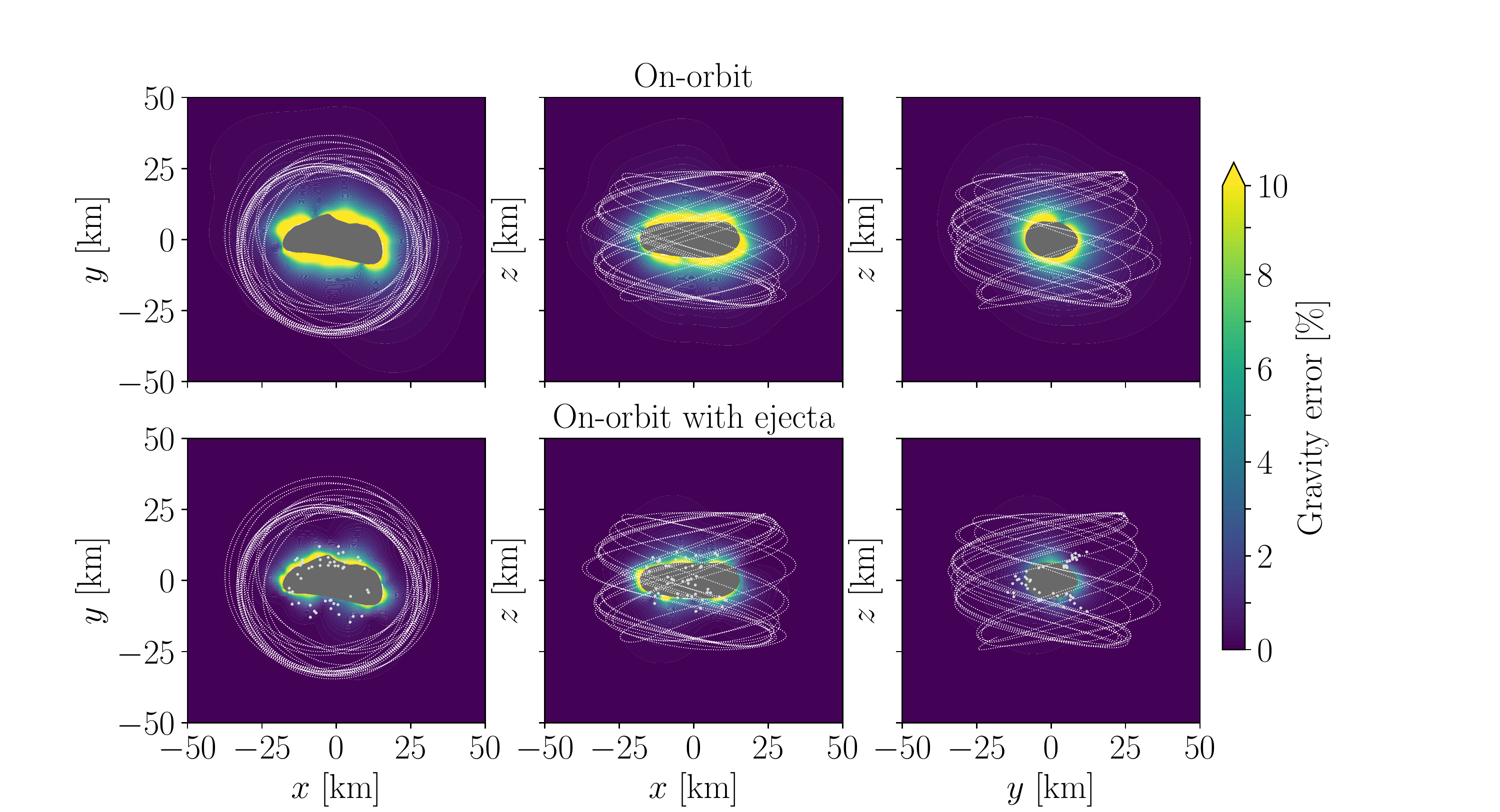}
	\end{center}
	\vspace{0cm}
	\caption{Gravity error maps with and without low altitude samples for scenario A1 and $n=100$.}	
	\label{fig:onlineejecta}
\end{figure}

Following paragraph~\ref{sec:orbittruth_accuracy}, the positive impact of including low altitude samples is to be analyzed. The same low altitude data of Fig.~\ref{fig:datasets} (\textit{right}) are used. In a real mission operation, these data samples are to be acquired while on the fly (which is out of the scope of this work). In order to account for that, a 5\% Gaussian error is added on each acceleration component. For scenario A1, 2D gravity error maps with and without low altitude samples (namely ejecta) are shown in Fig.~\ref{fig:onlineejecta}. It can be easily observed that these additional samples help to reduce significantly the error in the low altitude domain. The fact that the ejecta data is generated randomly but concentrated on specific spatial regions can be deduced from the $yz$ map. Table~\ref{tab:ejectaComparison} compares the global gravity error with and without ejecta for the on-orbit scenarios. For all the simulated scenarios (both with the complete and static trainings), the global gravity error is reduced significantly ($\approx50\%$).
\begin{table}[]  
	\centering
	\begin{tabular}{|c|cc|cc|}
		\cline{2-5}
		\multicolumn{1}{l}{} &  \multicolumn{2}{|c|}{Orbit} & \multicolumn{2}{|c|}{Orbit with ejecta} \\
		\hline
		Scen. &  $\mu_M$ & $\mu_M,\mathbf{r}_M$ & $\mu_M$ & $\mu_M,\mathbf{r}_M$\\
		\hline
		A1 & 3.674\% & 2.561\% & 1.424\% & 0.956\%\\
		\hline
		A2 & 4.129~\% & 3.614\% & 1.709\% & 1.572\%\\
		\hline  
		B1 & 3.591\% & 3.040\% & 1.542\% & 1.461\%\\
		\hline  
		B2 & 3.367\% & 2.661\% & 1.681\% & 1.624\%\\          
		\hline
	\end{tabular}
	\caption{Global gravity percent error with and without low altitude samples for $n=100$.}
	\label{tab:ejectaComparison}
\end{table}

\subsubsection{Computational efficiency}

Let conclude this subsection by evaluating the simultaneous navigation and gravity estimation computational efficiency. This paper simulations are executed in a M1 Max (2.06-3.22 GHz) processor using a single thread. Recall that the algorithms are coded in C++ under the Basilisk simulation framework (see Fig.~\ref{fig:BSK_diagram}). The execution times of the DMC-UKF step and each mascon optimization are reported in Table~\ref{tab:computational_times}. The number of masses ranges from 100 to 1000. The DMC-UKF execution is in the order of milliseconds or less. The mascon training ranges between 1 to 10 seconds for the static setup and 10-120 for the complete setup. Since the time span between measurements is $60~\text{s}$ and one orbit lasts $\approx 15~\text{h}$, the computation times show promise in terms of on-board autonomy.
\begin{table}[]  
	\centering
	\begin{tabular}{|c|c|c|c|c|}
		\cline{2-5}
		\multicolumn{1}{l}{} & \multicolumn{2}{|c|}{DMC-UKF step} & \multicolumn{2}{|c|}{Mascon training}\\
		\hline
		$n$ [-] & Mean $[\text{ms}]$ & Max. $[\text{ms}]$ & Mean $[\text{s}]$& Max. $[\text{s}]$\\
		\hline
		$100$ (M) & 0.289 & 0.779 & 1.099 & 1.146\\ 
		$500$ (M) & 1.194 & 2.742 & 5.890 & 6.091\\  
		$1000$ (M) & 2.327 & 3.648 & 12.53& 12.69\\
		\hline  
	    $100$ (MP) & 0.283 & 0.542 & 10.91 & 10.94\\ 
		$500$ (MP) & 1.192 & 2.461 & 56.56 & 56.96\\           
		$1000$ (MP)& 2.327 & 4.008 & 112.5 & 112.6\\               
		\hline
	\end{tabular}
	\caption{Computational times of simultaneous navigation and gravity estimation for several models. M $\equiv$ masses fit; MP $\equiv$ masses and positions fit.}
	\label{tab:computational_times}
\end{table}

\subsection{Propagation analysis}  

The intended purpose of the fitted mascon models is to serve as propagators for subsequent mission phases. Consequently, a propagation analysis is done in this subsection. The orbits to propagate vary their initial semi-major axis $a_0$ and inclination $i_0$ being $\{e_0=0.001,\Omega_0=48.2^{\circ},\omega_0=347.8^{\circ},\nu_0=85.3^{\circ}\}$ fixed. The analyzed set of inclinations is $i_0=\{0^{\circ},45^{\circ},90^{\circ},180^{\circ}\}$. The semi-major axis is varied between 28 to $46~\text{km}$ which ensures the orbit stability for the propagation duration of $12~\text{h}$. The propagation accuracy is measured by the position RMSE evaluated each $60~\text{s}$.

\begin{figure}[htbp] 
	\begin{center}
		\includegraphics[width=12cm,height=12cm,keepaspectratio,trim={0cm 0.05cm 0 0.25cm},clip]{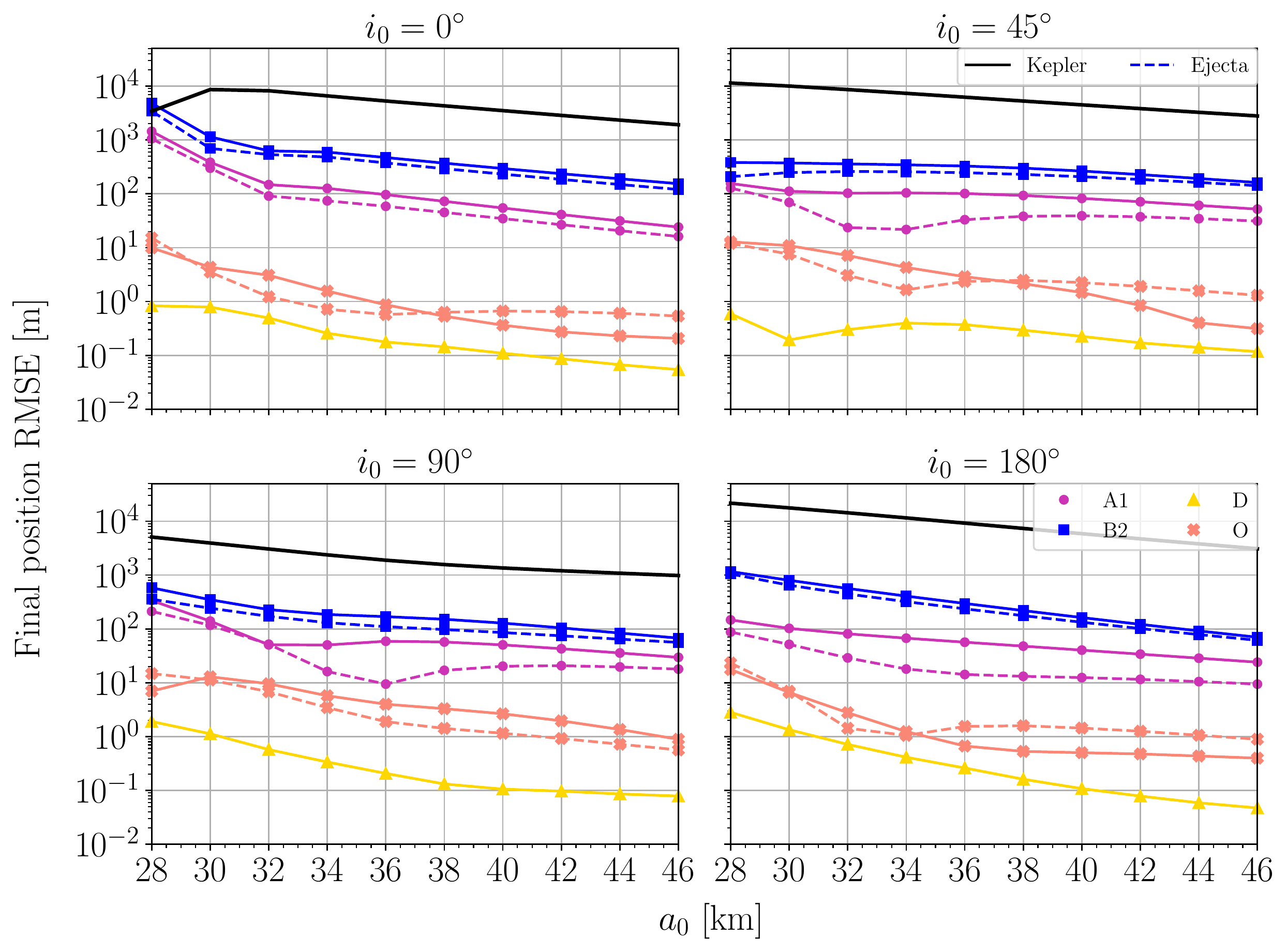}
	\end{center}
	\vspace{0cm}
	\caption{Final position RMSE after $12~\text{h}$ for different initial orbits with $n=1000$ mascon models.}	
	\label{fig:finalRMSE}
\end{figure}

The models under consideration are the ones that fit the complete mascon distribution. Amongst them, the analyzed models correspond to A1 and B2 scenarios (simultaneous navigation and gravity estimation), on-orbit and dense datasets (true data). The fits with low altitude samples are also considered. For these scenarios, the final position RMSE is shown in Fig.~\ref{fig:finalRMSE}. For the most severe scenario B2, the expected error ranges between $0.1$-$1~\text{km}$ in the majority of cases. For the less stringent scenario A1, with the low altitude samples, the error can be typically drive down to $10$-$100~\text{m}$. This may preclude the use of the fitted on-orbit mascon models in long-term propagations. However, by training a truth on-orbit dataset, the errors are reduced to $1$-$10~\text{m}$. This demonstrates that the on-orbit DMC-UKF dataset can be possibly improved with precise orbit determination. Finally, the mascon distribution trained with a dense dataset achieves a high level of accuracy ($0.1$-$1~\text{m}$ RMSE). This validates the effectiveness of the mascon model for small body propagations.

The results in Fig.~\ref{fig:finalRMSE} are shown for $n=1000$. Although the number of masses seems to not impact significantly the gravity errors (recall Fig.~\ref{fig:offlinenM} and Fig.~\ref{fig:onlinenM}), high $n$ values may be more robust to evaluation mismatches. This refers to the fact that in a propagation, the evaluation point diverges from the truth one over time. This intuition is confirmed by Fig.~\ref{fig:offlinenM} where for a specific propagation, $n$ is varied. It can be observed that $n=1000$ provides the best results in the majority of cases. Although it is concluded that the on-orbit mascon fits are not suitable for long term propagations, they may be useful for short-term ones. For example, after $2~\text{h}$, scenario B2 error is $\approx 20~\text{m}$ which greatly improves the Keplerian error of $\approx 300~\text{m}$. This could be of use for navigation or control purposes. 
\begin{figure}[htbp] 
	\begin{center}
		\includegraphics[width=8cm,height=8cm,keepaspectratio,trim={0cm 0.05cm 0 0.25cm},clip]{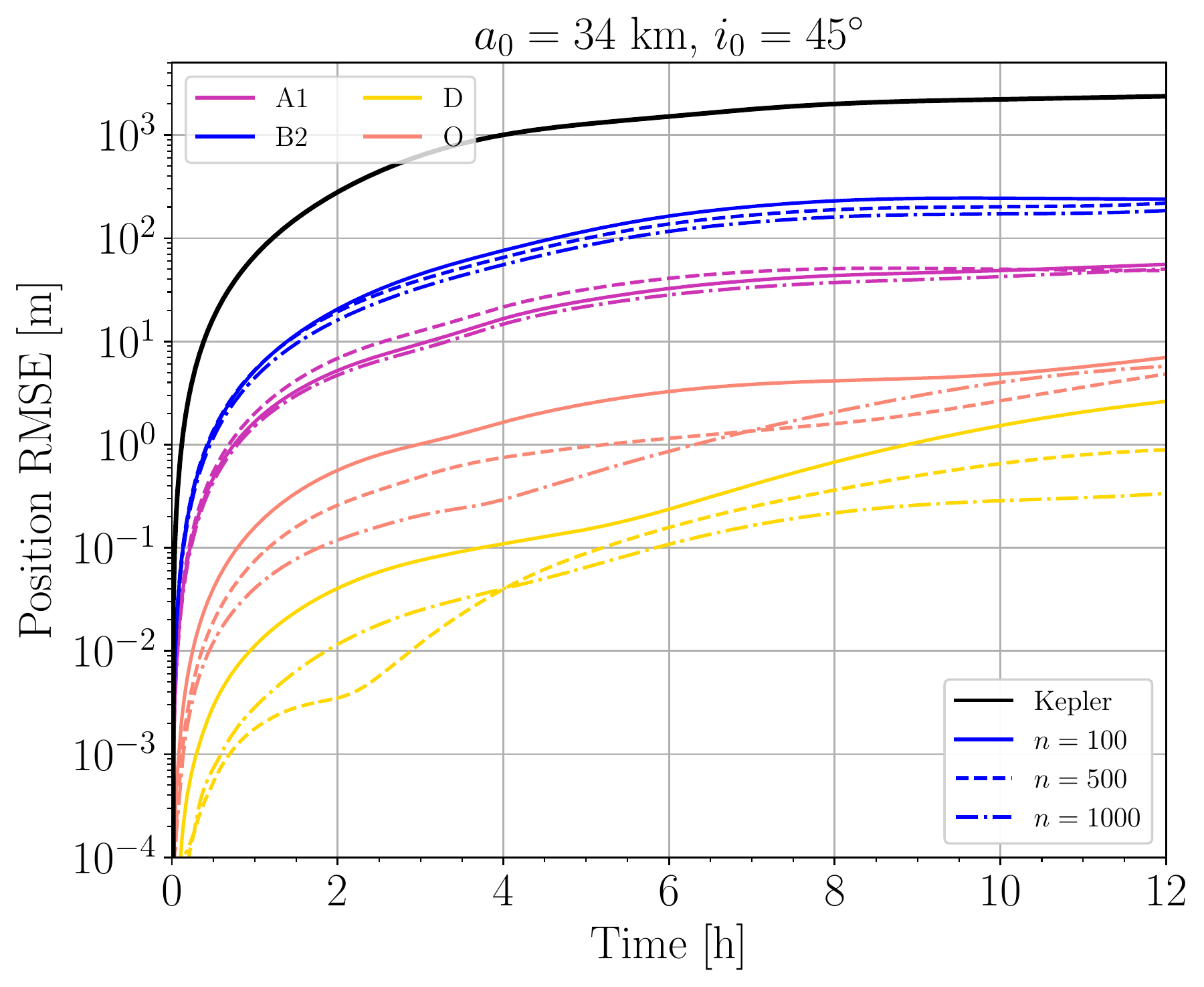}
	\end{center}
	\vspace{0cm}
	\caption{Position RMSE evolution with respect to number of masses.}	
	\label{fig:RMSEnM}
\end{figure} 

Finally, the computational efficiency of the mascon propagations is evaluated in Table~\ref{tab:computational_times_prop}. It can be easily observed that mascon models are competitive in terms of propagation speed (without any parallelization). The computational cost is only five times higher than a single Keplerian model for the most heavy model. At the same time, that model is 25 times more efficient than the polyhedron model. 
\begin{table}[]  
	\centering
	\begin{tabular}{|c|c|c|}
		\hline
		Model & Mean $[\text{s}]$ & Max. $[\text{s}]$\\
		\hline
		Kepler & 0.546 & 0.555 \\
		\hline
		Poly. 7790 faces & 68.11 & 68.31\\ 
		\hline
		Mascon $n=100$ & 0.779 & 0.792\\  
		Mascon $n=500$ & 1.668 & 1.687\\
		Mascon $n=1000$ & 2.734 & 2.801\\            
		\hline
	\end{tabular}
	\caption{Computational times of Kepler, polyhedron and mascon models for $12~\text{h}$ orbit propagation.}
	\label{tab:computational_times_prop}
\end{table}

\section{Conclusions}\label{sec:conclusions}

This manuscript presents a simultaneous navigation and mascon gravity estimation strategy around a small body. The scheme fuses the concept of dynamical model compensation with a mascon gravity optimizer. The dynamical model compensated unscented Kalman filter generates an on-orbit position-unmodeled acceleration dataset. The measurements are obtained by an on-board camera tracking surface landmarks. Then, the mascon distribution that best fits the on-orbit dataset is found. The mascon optimizer uses Adam gradient descent and ensures physical constraints through a projection step. Numerical simulations around asteroid Eros validate the proposed methodology. Even under different scenario conditions, the trained mascon models always improve the Keplerian model in the analyzed altitude regime. A half day propagation analysis reveals that the trained mascon distribution with the most severe conditions (measurement outages and landmarks uncertainty) yields $100$-$1000~\text{m}$ root mean square error. Still, there is room for future improvement since training under the truth on-orbit dataset reduces the propagation error to $1$-$10~\text{m}$.

Future work may focus on reducing the accuracy gap between the realistic on-orbit scenario and its equivalent truth dataset. To this end, several ideas has been preliminary tested in this paper. In particular, scenarios without lighting constraint (thus no measurements outages) and the addition of low altitude samples have a positive effect in the mascon accuracy. As an example, the aforementioned propagation errors are $10$-$100~\text{m}$ under continuous measurements. In the future, it is planned to deep into these ideas. For example, measurement outages can be precluded by a multi-satellite configuration which enables the use of inter-satellite ranging. Apart from the signal continuity, relative ranging is a highly accurate measurement that may facilitate precise orbit determination. The acquisition process of low altitude samples is also to be further explored. To this end, ejecta trajectories and their tracking by a specific sensor should be realistically simulated.

\section*{Acknowledgements}

This work is part of a project that has received funding from the European Union's Horizon 2020 research and innovation programme under the Marie Skłodowska-Curie grant agreement No. 101025257.

\bibliography{references}

\appendix

\section{Loss function gradient}\label{app:loss}

This appendix provides the expression of the loss function (see Eq.~\eqref{eq:loss}) gradient. The chain rule is exploited to separate the model derivatives from the loss choice.

\subsection{Mascon model Jacobian}

The Jacobian of the mascon model, for a dataset as Eq.~\eqref{eq:dataset}, has the following structure
\begin{equation}
	\mathbf{J}_{M}=[\mathbf{J}_{\mu_M}\>\> \mathbf{J}_{\mathbf{r}_M}],\label{eq:Jacobian}
\end{equation}
where $\mathbf{J}_{\mu_M}$ is Jacobian of the masses and $\mathbf{J}_{\mathbf{r}_M}$ refers to the spatial distribution. These can be expanded as 
\begin{equation}
	\mathbf{J}_{\mu_M}=2\mathbf{A}_{M}\text{diag}(\sqrt{\pmb{\mu}}_{\mathbf{S}_M}),\quad
	\mathbf{J}_{\mathbf{r}_M}=\begin{bmatrix}
		\dfrac{\partial{\mathbf{a}_{M}}}{\partial\mathbf{r}_{M_{1}}}\biggr\rvert_{\mathbf{r}_1} & \hdots & \dfrac{\partial{\mathbf{a}_{M}}}{\partial\mathbf{r}_{M_{n}}}\biggr\rvert_{\mathbf{r}_1}\\
		\vdots & \ddots & \vdots\\
		\dfrac{\partial{\mathbf{a}_{M}}}{\partial\mathbf{r}_{M_{1}}}\biggr\rvert_{\mathbf{r}_m} & \hdots & \dfrac{\partial{\mathbf{a}_{M}}}{\partial\mathbf{r}_{M_{n}}}\biggr\rvert_{\mathbf{r}_m}\\
	\end{bmatrix},
\end{equation}
where $\text{diag}()$ indicates a diagonal matrix. The Jacobian $\mathbf{J}_{\mathbf{r}_M}$ elements are
\begin{equation}
	\dfrac{\partial{\mathbf{a}_{M}}}{\partial\mathbf{r}_{M_k}}\biggr\rvert_{\mathbf{r}_j}=\mu_{M_k}\left(\dfrac{1}{\Delta r^3_{jk}}\mathbf{I}-\dfrac{3\Delta\mathbf{r}_{jk}\Delta\mathbf{r}^T_{jk}}{\Delta r^5_{jk}}\right).
\end{equation}
Note that Eq.~\eqref{eq:Jacobian} structure is flexible. For example, if is decided to just fit $\sqrt{\mu_{M_k}}$ being $\mathbf{r}_{M_k}$ fixed, the spatial distribution Jacobian $\mathbf{J}_{\mathbf{r}_M}$ vanishes from the expression. Then, the Jacobian operations are very efficient since the matrix $\mathbf{A}_M$ needs only to be computed once (as it only depends on $\mathbf{r}_M$ which is fixed in this case).  

\subsection{Gradient with respect to acceleration}

The loss function and the mascon model are related by the gravity acceleration. The loss function gradient with respect to the mascon acceleration prediction is to be computed. Since the loss (see Eq.~\eqref{eq:loss}) is chosen to be the mean squared percent error, this gradient is
\begin{equation}
\nabla L_{\mathbf{a}_M}=\frac{1}{m}\begin{bmatrix}
	2(\mathbf{a}_M(\mathbf{r}_1)-\mathbf{a}_1)/a^2_1\\
	\vdots\\
	2(\mathbf{a}_M(\mathbf{r}_m)-\mathbf{a}_m)/a^2_m]\\
	\end{bmatrix}.
\end{equation} 
Any other choice of loss function is admitted as long as its gradient with respect to the acceleration mascon prediction can be obtained.

\subsection{Chain rule}

By applying the chain rule, the loss function gradient with respect to the mascon distribution variables is
\begin{equation}
\nabla L = \begin{bmatrix}
\mathbf{1}^T\left(\nabla L_{\mathbf{a}_M}\oplus\dfrac{d\mathbf{a}_M}{d\sqrt{\mu}_1}\right)\\
\vdots\\
\mathbf{1}^T\left(\nabla L_{\mathbf{a}_M}\oplus\dfrac{d\mathbf{a}_M}{dx_{M_n}}\right)\\
\mathbf{1}^T\left(\nabla L_{\mathbf{a}_M}\oplus\dfrac{d\mathbf{a}_M}{dy_{M_n}}\right)\\
\mathbf{1}^T\left(\nabla L_{\mathbf{a}_M}\oplus\dfrac{d\mathbf{a}_M}{dz_{M_n}}\right)\\
\end{bmatrix},\label{eq:loss_gradient}
\end{equation}
where $\mathbf{1}$ is a vector of ones that helps expressing the summation of the right side vector components. Although the expression may seem complex, the vector within the parentheses is nothing more than the element-wise product of $\nabla L_{\mathbf{a}_M}$ with each column of the mascon model Jacobian $\mathbf{J}_M$.

\section{Static position determination}\label{app:camera}

This appendix provides a fast analysis tool for the landmarks-based navigation errors. To this end, a static landmarks-based position determination algorithm is described. Then, the position error distribution with respect to the camera focal length is analyzed.

\subsection{Landmark-based position determination}\label{app:camera_position}

A position fix can be approximately determined by solving the underlying landmarks-based navigation geometrical problem. If no errors (attitude, landmarks position, pixelation) are present, the lines between a surface landmark and its projection on the image plane uniquely intersect in the camera aperture centre (assumed as spacecraft center of mass for simplicity). This is deduced from Fig.~\ref{fig:camerapinhole_model}. When errors are present, the intersection is not unique. Alternatively, the position fix can be approximated by the closest point with respect to all lines in terms of the mean squared distance. In order to describe that logic, let first detail how the line between the landmark and its pixel can be constructed:
\begin{enumerate}
	\item{Transform the center pixel to 2D image coordinates: $(u_{l},v_{l})=w_p(p_{x_l},p_{y_l})$.}
	\item{Compute the line-of-sight vector $\mathbf{w}^{C}_l$ of the landmark from the spacecraft:
	\begin{equation*} \mathbf{w}^{C}_l=\frac{1}{\sqrt{u^2_{l}+v^2_{l}+1}}\begin{bmatrix}u_{l}\\v_{l}\\1\end{bmatrix}.
	\end{equation*}}
	\item{Project the line-of-sight vector into the rotating small body centred frame: $\mathbf{w}_l=(\mathbf{R}^{C}_{A})^T\mathbf{w}_l^{C}$.}
	\item{Construct the parametric equation of the line: $s\mathbf{w}_l=\mathbf{s}-\mathbf{r}_{l}$.}
\end{enumerate}
Note that $s$ is the independent variable and $\mathbf{s}$ is the set of points sweeping along the line. The distance $d_l$ of a point $\mathbf{r}'$ to a line can be computed as
\begin{equation}
	d_l = \lVert(\mathbf{r}'-\mathbf{r}_{l})\times\mathbf{w}_l \rVert_2.
\end{equation}
Recall that $\mathbf{w}_l$ is a unit vector. Then, the goal is to find the point $\hat{\mathbf{r}}$ that minimizes the sum of the squared distances $D$ for a number $n_L$ of landmarks
\begin{equation}
\hat{\mathbf{r}}\approx\underset{\mathbf{r}'}{\text{argmin}}\>\>D=\underset{\mathbf{r}'}{\text{argmin}}\>\>\sum^{n_L}_{l=1}d^2_l,\label{eq:mindistances_problem}
\end{equation}
where the sum of the distances can be expanded as 
\begin{equation}
	\begin{aligned}
			\sum^{n_L}_{l=1}d^2_l&=\sum^{n_L}_{l=1}
			[(\mathbf{r}'-\mathbf{r}_{l})\times\mathbf{w}_l]^T	[(\mathbf{r}'-\mathbf{r}_{l})\times\mathbf{w}_l]\\
			&=\sum^{n_L}_{l=1}(\mathbf{r}'-\mathbf{r}_{l})^T(\mathbf{r}'-\mathbf{r}_{l})-[(\mathbf{r}'-\mathbf{r}_{l})\cdot\mathbf{w}_l]^T[(\mathbf{r}'-\mathbf{r}_{l})\cdot\mathbf{w}_l].\label{eq:sum_distances}
	\end{aligned}
\end{equation}
Note that the vector identity $(\mathbf{a}\times\mathbf{b})^T(\mathbf{a}\times\mathbf{b})=(\mathbf{a}^T\mathbf{a})(\mathbf{b}^T\mathbf{b})-(\mathbf{a}^T\mathbf{b})(\mathbf{b}^T\mathbf{a})$ is used.
The spacecraft position estimate $\hat{\mathbf{r}}$, can be computed by taking the derivative of $D$ with respect to $\mathbf{r}'$ and equaling the resulting expression to zero
\begin{equation}
\frac{dD}{d\mathbf{r}'}\biggr\rvert_{\mathbf{r}'=\hat{\mathbf{r}}}=\sum\limits^{n_L}_{l=1}2(\hat{\mathbf{r}}-\mathbf{r}_{l})-2[(\hat{\mathbf{r}}-\mathbf{r}_{l})\cdot\mathbf{w}_l]\mathbf{w}_l=\mathbf{0}.
\end{equation}
After rearranging terms, $\hat{\mathbf{r}}$ can be cleared as
\begin{equation}
\left(n_L\mathbf{I}-\sum\limits^{n_L}_{l=1}\mathbf{w}_l(\mathbf{w}_l)^T\right)\hat{\mathbf{r}}=\sum\limits^{n_L}_{l=1}\mathbf{r}_{l} - (\mathbf{r}_{l}\cdot\mathbf{w}_l )\mathbf{w}_l,\label{eq:position_determination}
\end{equation}
which is a 3x3 system of linear equations that can be solved by inverting the left-side matrix. Note that $\mathbf{I}$ denotes the identity matrix. There are two singular cases for Eq.~\eqref{eq:position_determination} system. The first one is $n_L<2$ where it is evident that there are not enough lines to determine an intersection. The other case is when all the lines are parallel which may occur if the visible landmarks are too close (thus projecting to the same pixel on the image).

\subsection{Position error with respect to focal length}

\begin{figure}[htbp] 
	\begin{center}
		\includegraphics[width=7cm,height=7cm,keepaspectratio,trim={0cm 0.25cm 0 0.25cm},clip]{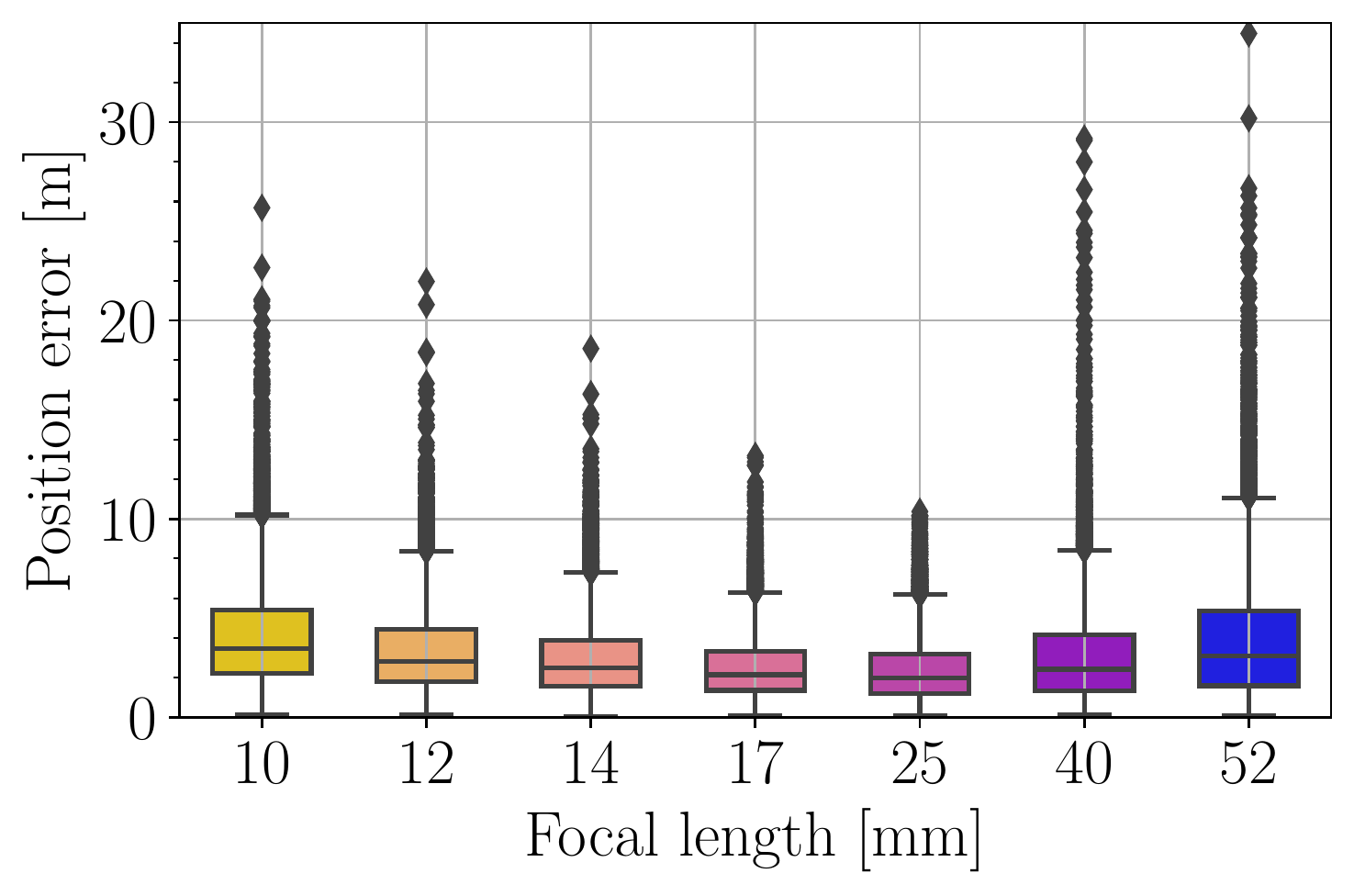}
	\end{center}
	\vspace{0cm}
	\caption{Distribution of position errors with respect to focal length.}	
	\label{fig:errCamera}
\end{figure}

The static position determination algorithm of~\ref{app:camera_position} allows a preliminary analysis of navigation errors. The main advantage is that this analysis is purely geometrical and free of dynamics. In the case under consideration, it is of interest to find a suitable camera focal length $f$ (which controls how much is visible and its resolution) for subsection~\ref{sec:results:simultaneous}. To this end, several focal lengths ranging the diagonal FOV evenly between $23^{\circ}$ and $94^{\circ}$ are analyzed for the orbit under consideration. The distribution of position errors is shown in Fig.~\ref{fig:errCamera}. The simulation does not include the lighting constraint in order to avoid an excessive number of outliers. After examining Fig.~\ref{fig:errCamera}, the value $f=25~\text{mm}$ is used in this manuscript.

%Since the position is determined as the solution of a least-squares problem (see Eq.~\eqref{eq:mindistances_problem}), its uncertainty can be approximated. In particular, the estimated position covariance matrix is
%\begin{equation}
%\pmb{\Sigma}_{\mathbf{z}\mathbf{z}}\approx\chi^2(\mathbf{J}_D\mathbf{J}_D^T)^{-1},\label{eq:covariance_determination}
%\end{equation}
%where $\chi^2$ is the residuals chi-squared statistic and $\mathbf{J}_D$ is a Jacobian matrix composed of each pseudoline distance $d_j$ derivative with respect to $\mathbf{r}^*$
%\begin{equation} \chi^2=\frac{D(\mathbf{r})}{3n_L-3},\quad \mathbf{J}_D=[(\mathbf{w}_1)^{\times},\hdots,(\mathbf{w}_{n_L})^{\times}],
%\end{equation}   
%being $(\mathbf{w}_l)^{\times}$ the cross-product matrix of each line-of-sight vector. Finally, the DMC-UKF measurement input is the following Gaussian distribution
%\begin{equation}
%\mathbf{r}\sim N_3(\hat{\mathbf{r}},\pmb{\Sigma}_{\mathbf{z}\mathbf{z}}).\label{eq:position_determination_gaussian}
%\end{equation}

%% \label{}

\end{document}